# The symmetries of the Fokker - Planck equation in three dimensions

*Igor A. Tanski*
*tanski@protek.ru*

ZAO CV Protek

*ABSTRACT*

We calculate all point symmetries of the Fokker - Planck equation in three-dimensional Euclidean space. General expression of symmetry group action on arbitrary solution of Fokker - Planck equation is presented.

## 1. The symmetries of the Fokker - Planck equation in three dimensions

The object of our considerations is a special case of Fokker - Planck equation, which describes evolution of 3D continuum of non-interacting particles imbedded in a dense medium without outer forces. The interaction between particles and medium causes combined diffusion in physical space and velocities space. The only force, which acts on particles, is damping force proportional to velocity.

The 3D variant of this equation was investigated in our work [1]. In this work fundamental solution of 3D equation was obtained by means of Fourier transform.

The 1D and 2D variants of this equation were investigated in our works [2-3]. All point symmetries of the Fokker - Planck equation in one-dimensional Euclidean space were calculated.

In present work we continue this investigation for more complex 3D equation.

The Fokker - Planck equation in three dimensions is

$$\frac{\partial n}{\partial t} + u\frac{\partial n}{\partial x} + v\frac{\partial n}{\partial y} + w\frac{\partial n}{\partial z} - au\frac{\partial n}{\partial u} - av\frac{\partial n}{\partial v} - aw\frac{\partial n}{\partial z} - 3an - k\left(\frac{\partial^2 n}{\partial u^2} + \frac{\partial^2 n}{\partial v^2} + \frac{\partial^2 n}{\partial w^2}\right) = 0. \qquad (1)$$

where

$n = n(t, x, y, z, u, v, w)$ - density;

$t$ - time variable;

$x, y, z$ - space coordinates;

$u, v, w$ - velocity;

$a$ - coefficient of damping;

$k$ - coefficient of diffusion.

The list of symmetries of the Fokker - Planck equation in three dimensions follows. The calculations of symmetries are rather awkward. They are carried out to APPENDIX 1.

Instead of classic "$\xi - \phi$" notation we use another ("$\delta$") notation. This notation was presented in our work [2].

Addition of arbitrary solution

$$\mathbf{v}_1 = A\frac{\partial}{\partial n}; \qquad (2)$$

where $A$ is arbitrary solution of the (1) equation.

Scaling of density

$$\mathbf{v}_2 = n\,\frac{\partial}{\partial n}\,. \tag{3}$$

The reason of symmetries (2-3) existence is linearity of PDE (1).

Time shift

$$\mathbf{v}_3 = \frac{\partial}{\partial t}\,. \tag{4}$$

Space translations

$$\mathbf{v}_4 = \frac{\partial}{\partial x}\,; \quad \mathbf{v}_5 = \frac{\partial}{\partial y}\,; \quad \mathbf{v}_6 = \frac{\partial}{\partial z}\,. \tag{5}$$

Space rotations

$$\mathbf{v}_7 = w\,\frac{\partial}{\partial v} - v\,\frac{\partial}{\partial w} + z\,\frac{\partial}{\partial y} - y\,\frac{\partial}{\partial z}\,; \; \mathbf{v}_8 = u\,\frac{\partial}{\partial w} - w\,\frac{\partial}{\partial u} + x\,\frac{\partial}{\partial z} - z\,\frac{\partial}{\partial x}\,; \; \mathbf{v}_9 = v\,\frac{\partial}{\partial u} - u\,\frac{\partial}{\partial v} + y\,\frac{\partial}{\partial x} - x\,\frac{\partial}{\partial y}\,. \tag{6}$$

Transformations (5) and (6) build three-dimensional Euclidean movements group.

Extended Galilean transformations, which besides time and space coordinates affect the density

$$\mathbf{v}_{10} = \frac{\partial}{\partial u} + t\,\frac{\partial}{\partial x} - \frac{an}{2k}\,(ax+u)\,\frac{\partial}{\partial n}\,; \quad \mathbf{v}_{11} = \frac{\partial}{\partial v} + t\,\frac{\partial}{\partial y} - \frac{an}{2k}\,(ay+v)\,\frac{\partial}{\partial n}\,; \tag{7}$$

$$\mathbf{v}_{12} = \frac{\partial}{\partial w} + t\,\frac{\partial}{\partial z} - \frac{an}{2k}\,(az+w)\,\frac{\partial}{\partial n}\,.$$

Negative exponent transformations - they affect time and space coordinates, contain time-dependent common multiplier (negative exponent). They do not affect density.

$$\mathbf{v}_{13} = e^{-at}\left(-a\,\frac{\partial}{\partial u} + \frac{\partial}{\partial x}\right); \; \mathbf{v}_{14} = e^{-at}\left(-a\,\frac{\partial}{\partial v} + \frac{\partial}{\partial y}\right); \; \mathbf{v}_{15} = e^{-at}\left(-a\,\frac{\partial}{\partial w} + \frac{\partial}{\partial z}\right); \tag{8}$$

Positive exponent transforms - they affect time, space and density, contain time-dependent common multiplier (positive exponent).

$$\mathbf{v}_{16} = e^{at}\left(a\,\frac{\partial}{\partial u} + \frac{\partial}{\partial x} - \frac{a^2}{k}\,nv\,\frac{\partial}{\partial n}\right); \; \mathbf{v}_{17} = e^{at}\left(a\,\frac{\partial}{\partial v} + \frac{\partial}{\partial y} - \frac{a^2}{k}\,nv\,\frac{\partial}{\partial n}\right); \tag{9}$$

$$\mathbf{v}_{18} = e^{at}\left(a\,\frac{\partial}{\partial w} + \frac{\partial}{\partial z} - \frac{a^2}{k}\,nv\,\frac{\partial}{\partial n}\right);$$

One-parameter groups, generated by vector fields $\mathbf{v}_1 - \mathbf{v}_{18}$, are enumerated in the following list. The list contains images of the point $(n, t, x, y, z, u, v, w)$ by transformation $exp(\varepsilon \mathbf{v}_i)$

$G_1$:  $(n + \varepsilon A, t, x, y, z, u, v, w)$;

$G_2$:  $(e^\varepsilon n, t, x, y, z, u, v, w)$;

$G_3$:  $(n, t + \varepsilon, x, y, z, u, v, w)$;

$G_4$:  $(n, t, x + \varepsilon, y, z, u, v, w)$;

$G_5$:  $(n, t, x, y + \varepsilon, z, u, v, w)$;

$G_6$:  $(n, t, x, y, z + \varepsilon, u, v, w);$

$G_7$:  $(n, t, x, (\cos(\varepsilon)y + \sin(\varepsilon)z), (-\sin(\varepsilon)y + \cos(\varepsilon)z), u, (\cos(\varepsilon)v + \sin(\varepsilon)w), (-\sin(\varepsilon)v + \cos(\varepsilon)w));$

$G_8$:  $(n, t, (\cos(\varepsilon)x - \sin(\varepsilon)z), y, (\sin(\varepsilon)x + \cos(\varepsilon)z), (\cos(\varepsilon)u - \sin(\varepsilon)w), v, (\sin(\varepsilon)u + \cos(\varepsilon)w));$

$G_9$:  $(n, t, (\cos(\varepsilon)x + \sin(\varepsilon)y), (-\sin(\varepsilon)x + \cos(\varepsilon)y), z, (\cos(\varepsilon)u + \sin(\varepsilon)v), (-\sin(\varepsilon)u + \cos(\varepsilon)v), w);$

$G_{10}$:  $(\exp\left[-\dfrac{a}{2k}\left(\varepsilon(ax + u) + \dfrac{1}{2}\varepsilon^2(at + 1)\right)\right] n, t, x + \varepsilon t, y, z, u + \varepsilon, v, w);$  (10)

$G_{11}$:  $(\exp\left[-\dfrac{a}{2k}\left(\varepsilon(ay + v) + \dfrac{1}{2}\varepsilon^2(at + 1)\right)\right] n, t, x, y + \varepsilon t, z, u, v + \varepsilon, w);$

$G_{12}$:  $(\exp\left[-\dfrac{a}{2k}\left(\varepsilon(az + w) + \dfrac{1}{2}\varepsilon^2(at + 1)\right)\right] n, t, x, y, z + \varepsilon t, u, v, w + \varepsilon);$

$G_{13}$:  $(n, t, x + \varepsilon e^{-at}, y, z, u - \varepsilon a\ e^{-at}, v, w);$

$G_{14}$:  $(n, t, x, y + \varepsilon e^{-at}, z, u, v - \varepsilon a\ e^{-at}, w);$

$G_{15}$:  $(n, t, x, y, z + \varepsilon e^{-at}, u, v, w - \varepsilon a\ e^{-at});$

$G_{16}$:  $(n \exp\left[-\dfrac{a^2}{k} e^{at}\left(\varepsilon u + \dfrac{1}{2}\varepsilon^2 a\ e^{at}\right)\right], t, x + \varepsilon e^{at}, y, z, u + \varepsilon a\ e^{at}, v, w);$

$G_{17}$:  $(n \exp\left[-\dfrac{a^2}{k} e^{at}\left(\varepsilon v + \dfrac{1}{2}\varepsilon^2 a\ e^{at}\right)\right], t, x, y + \varepsilon e^{at}, z, u, v + \varepsilon a\ e^{at}, w);$

$G_{18}$:  $(n \exp\left[-\dfrac{a^2}{k} e^{at}\left(\varepsilon w + \dfrac{1}{2}\varepsilon^2 a\ e^{at}\right)\right], t, x, y, z + \varepsilon e^{at}, u, v, w + \varepsilon a\ e^{at}).$

For relatively nontrivial integration of $G_{10} - G_{12}$, $G_{16} - G_{18}$. we refer to [2] (APPENDIX 3-4).

The fact, that $G_i$ are symmetries of PDE (1) means, that if $f(t, x, y, u, v)$ is arbitrary solution of (1), the functions

$u^{(1)}$:  $f(t, x, y, z, u, v, w) + \varepsilon A(t, x, y, z, u, v, w);$

$u^{(2)}$:  $e^{\varepsilon} f(t, x, y, z, u, v, w);$

$u^{(3)}$:  $f(t - \varepsilon, x, y, z, u, v, w);$

$u^{(4)}$:  $f(t, x - \varepsilon, y, z, u, v, w);$

$u^{(5)}$:  $f(t, x, y - \varepsilon, z, u, v, w);$

$u^{(6)}$:  $f(t, x, y, z - \varepsilon, u, v, w);$

$u^{(7)}$:  $f(t, x, (\cos(\varepsilon)y - \sin(\varepsilon)z), (\sin(\varepsilon)y + \cos(\varepsilon)z), u, (\cos(\varepsilon)v - \sin(\varepsilon)w), (\sin(\varepsilon)v + \cos(\varepsilon)w));$

$u^{(8)}$:     $f(t, (\cos(\varepsilon)x + \sin(\varepsilon)z), y, (-\sin(\varepsilon)x + \cos(\varepsilon)z), (\cos(\varepsilon)u + \sin(\varepsilon)w), v, (-\sin(\varepsilon)u + \cos(\varepsilon)w));$

$u^{(9)}$:     $f(t, (\cos(\varepsilon)x - \sin(\varepsilon)y), (\sin(\varepsilon)x + \cos(\varepsilon)y), z, (\cos(\varepsilon)u - \sin(\varepsilon)v), (\sin(\varepsilon)u + \cos(\varepsilon)v), w);$ (11)

$u^{(10)}$:     $\exp\left[-\dfrac{a}{2k}\left(\varepsilon(ax + u) - \dfrac{1}{2}\varepsilon^2(at + 1)\right)\right]f(t, x - \varepsilon t, y, z, u - \varepsilon, v, w);$

$u^{(11)}$:     $\exp\left[-\dfrac{a}{2k}\left(\varepsilon(ay + v) - \dfrac{1}{2}\varepsilon^2(at + 1)\right)\right]f(t, x, y - \varepsilon t, z, u, v - \varepsilon, w);$

$u^{(12)}$:     $\exp\left[-\dfrac{a}{2k}\left(\varepsilon(az + w) - \dfrac{1}{2}\varepsilon^2(at + 1)\right)\right]f(t, x, y, z - \varepsilon t, u, v, w - \varepsilon);$

$u^{(13)}$:     $f(t, x - \varepsilon e^{-at}, y, z, u + \varepsilon a\ e^{-at}, v, w);$

$u^{(14)}$:     $f(t, x, y - \varepsilon e^{-at}, z, u, v + \varepsilon a\ e^{-at}, w);$

$u^{(15)}$:     $f(t, x, y, z - \varepsilon e^{-at}, u, v, w + \varepsilon a\ e^{-at});$

$u^{(16)}$:     $\exp\left[-\dfrac{a^2}{k}e^{at}\left(\varepsilon u - \dfrac{1}{2}\varepsilon^2 a\ e^{at}\right)\right]f(t, x - \varepsilon e^{at}, y, z, u - \varepsilon a e^{at}, v, w);$

$u^{(17)}$:     $\exp\left[-\dfrac{a^2}{k}e^{at}\left(\varepsilon v - \dfrac{1}{2}\varepsilon^2 a\ e^{at}\right)\right]f(t, x, y - \varepsilon e^{at}, z, u, v - \varepsilon a e^{at}, w);$

$u^{(18)}$:     $\exp\left[-\dfrac{a^2}{k}e^{at}\left(\varepsilon w - \dfrac{1}{2}\varepsilon^2 a\ e^{at}\right)\right]f(t, x, y, z - \varepsilon e^{at}, u, v, w - \varepsilon a e^{at}).$

where $\varepsilon$ - arbitrary real number , also are solutions of (1). Here $A$ is another arbitrary solution of (1).

We systematically replace "old coordinates" by their expressions through "new coordinates". Note, that due to these replacements terms with $\varepsilon^2$ in $u^{(10)}$ - $u^{(12)}$ and $u^{(16)}$ - $u^{(18)}$ change their signs.

We have trivial solution $n = e^{3at}$ at our disposal. If we act on this solution by transformations (10), we obtain 6 new solutions:

$$n = \exp\left[3at - \frac{a}{2k}\left(\varepsilon(ax + u) - \frac{1}{2}\varepsilon^2(at + 1)\right)\right]; \tag{19}$$

$$n = \exp\left[3at - \frac{a}{2k}\left(\varepsilon(ay + v) - \frac{1}{2}\varepsilon^2(at + 1)\right)\right]. \tag{20}$$

$$n = \exp\left[3at - \frac{a}{2k}\left(\varepsilon(az + w) - \frac{1}{2}\varepsilon^2(at + 1)\right)\right]; \tag{21}$$

$$n = \exp\left[3at - \frac{a^2}{k}e^{at}\left(\varepsilon u - \frac{1}{2}\varepsilon^2 a\ e^{at}\right)\right]. \tag{22}$$

$$n = \exp\left[3at - \frac{a^2}{k}e^{at}\left(\varepsilon v - \frac{1}{2}\varepsilon^2 a\ e^{at}\right)\right]. \tag{23}$$

$$n = \exp\left[3at - \frac{a^2}{k}\,e^{at}\left(\varepsilon w - \frac{1}{2}\,\varepsilon^2 a\,e^{at}\right)\right]. \tag{24}$$

General expression is

$$U = e^{\varepsilon_2}\exp\left[-\frac{a}{2k}\left(\varepsilon_{10}(a\bar{x}+\bar{u}) - \frac{1}{2}\,\varepsilon_{10}^2(at+1)\right)\right]\exp\left[-\frac{a}{2k}\left(\varepsilon_{11}(a\bar{y}+\bar{v}) - \frac{1}{2}\,\varepsilon_{11}^2(at+1)\right)\right]\exp\left[-\frac{a}{2k}\left(\varepsilon_{12}(a\bar{z}+\bar{w}) - \frac{1}{2}\,\varepsilon_{12}^2(at+1)\right)\right]\times \tag{25}$$

$$\times\exp\left[-\frac{a^2}{k}\,e^{at}\left(\varepsilon_{16}(\bar{u}-\varepsilon_{10}) - \frac{1}{2}\,\varepsilon_{16}^2 a\,e^{at}\right)\right]\exp\left[-\frac{a^2}{k}\,e^{at}\left(\varepsilon_{17}(\bar{v}-\varepsilon_{11}) - \frac{1}{2}\,\varepsilon_{17}^2 a\,e^{at}\right)\right]\exp\left[-\frac{a^2}{k}\,e^{at}\left(\varepsilon_{18}(\bar{w}-\varepsilon_{12}) - \frac{1}{2}\,\varepsilon_{18}^2 a\,e^{at}\right)\right]\times$$

$$\times f(t - \varepsilon_3, \bar{x} - \varepsilon_4 - \varepsilon_{10}t - \varepsilon_{13}e^{-at} - \varepsilon_{16}e^{at}, \bar{y} - \varepsilon_5 - \varepsilon_{11}t - \varepsilon_{14}e^{-at} - \varepsilon_{17}e^{at}, \bar{z} - \varepsilon_6 - \varepsilon_{12}t - \varepsilon_{15}e^{-at} - \varepsilon_{18}e^{at},$$

$$\bar{u} - \varepsilon_{10} + \varepsilon_{13}a\,e^{-at} - \varepsilon_{16}a\,e^{at}, \bar{v} - \varepsilon_{11} + \varepsilon_{14}a\,e^{-at} - \varepsilon_{17}a\,e^{at}, \bar{w} - \varepsilon_{12} + \varepsilon_{15}a\,e^{-at} - \varepsilon_{18}a\,e^{at}) + \varepsilon_1 A(t, x, y, z, u, v, w);$$

where

$$\bar{x} = \left(\cos(\varepsilon_8)\cos(\varepsilon_9)\right)x + \left(\cos(\varepsilon_8)\sin(\varepsilon_9)\right)y + (-\sin(\varepsilon_8))\,z; \tag{26}$$

$$\bar{y} = \left(\sin(\varepsilon_7)\sin(\varepsilon_8)\cos(\varepsilon_9) - \cos(\varepsilon_7)\sin(\varepsilon_9)\right)x + \left(\sin(\varepsilon_7)\sin(\varepsilon_8)\sin(\varepsilon_9) + \cos(\varepsilon_7)\cos(\varepsilon_9)\right)y + \tag{27}$$

$$+(\sin(\varepsilon_7)\cos(\varepsilon_8))\,z;$$

$$\bar{z} = \left(\sin(\varepsilon_7)\sin(\varepsilon_9) + \cos(\varepsilon_7)\sin(\varepsilon_8)\cos(\varepsilon_9)\right)x + \left(-\sin(\varepsilon_7)\cos(\varepsilon_9) + \cos(\varepsilon_7)\sin(\varepsilon_8)\sin(\varepsilon_9)\right)y + \tag{28}$$

$$+(\cos(\varepsilon_7)\cos(\varepsilon_8))\,z;$$

$$\bar{u} = \left(\cos(\varepsilon_8)\cos(\varepsilon_9)\right)u + \left(\cos(\varepsilon_8)\sin(\varepsilon_9)\right)v + (-\sin(\varepsilon_8))\,w; \tag{29}$$

$$\bar{v} = \left(\sin(\varepsilon_7)\sin(\varepsilon_8)\cos(\varepsilon_9) - \cos(\varepsilon_7)\sin(\varepsilon_9)\right)u + \left(\sin(\varepsilon_7)\sin(\varepsilon_8)\sin(\varepsilon_9) + \cos(\varepsilon_7)\cos(\varepsilon_9)\right)v + \tag{30}$$

$$+(\sin(\varepsilon_7)\cos(\varepsilon_8))\,w;$$

$$\bar{w} = \left(\sin(\varepsilon_7)\sin(\varepsilon_9) + \cos(\varepsilon_7)\sin(\varepsilon_8)\cos(\varepsilon_9)\right)u + \left(-\sin(\varepsilon_7)\cos(\varepsilon_9) + \cos(\varepsilon_7)\sin(\varepsilon_8)\sin(\varepsilon_9)\right)v + \tag{31}$$

$$+(\cos(\varepsilon_7)\cos(\varepsilon_8))\,w.$$

## DISCUSSION

Looking at the list of all point symmetries of the Fokker - Planck equation in two-dimensional Euclidean space, we see, that there is no simple way to get, for example, fundamental solution of PDE, using these symmetries. We have not at our disposal such an instrument, as scaling of independent variables $t, x, y, z, u, v, w$. The result (19-24) of action of symmetry group on trivial solution is not very interesting from physical point of view.

Indirect way of use of Galilean transformations (7) was demonstrated in [1]. The transformation was used for generalization of solution, which was obtained in the form of exponent of quadratic form of space coordinates and velocities with time dependent coefficients.

There is need of further investigations of Fokker - Planck equation and its set of symmetries, which may lead to another physically interesting results. We can follow the scheme of [8] : to consider invariant solutions for some one-parameter group, thus reduce the independent variables number. To find for obtained in such a way equation all point symmetries - and so long.

In the work [8] this scheme was represented for equations of elasticity and plasticity.

## ACKNOWLEDGMENTS

We wish to thank Jos A. M. Vermaseren from NIKHEF (the Dutch Institute for Nuclear and High-Energy Physics), for he made his symbolic computations program FORM release 3.1 available for download for non-commercial purposes (see [9]). This wonderful program makes difficult task of symmetries search more accessible.

___________________________

## REFERENCES

[1] Igor A. Tanski. Fundamental solution of Fokker - Planck equation. arXiv:nlin.SI/0407007 v1 4 Jul 2004

[2] Igor A. Tanski. The symmetries of the Fokker - Planck equation in two dimensions. arXiv:nlin.CD/0412075 v1 31 Dec 2004

[3] Igor A. Tanski. The symmetries of the Fokker - Planck equation in one dimension. arXiv:nlin.PS/0501008 v1 3 Jan 2005

[4] L. P. Eisenhart, Continuous Groups of Transformations, Princeton University Press, Princeton, 1933.

[5] L. V. Ovsyannikov, Group analysis of differential equations, Moscow, Nauka, 1978.

[6] Peter J. Olver, Applications of Lie groups to differential equations. Springer-Verlag, New York, 1986.

[7] N. H. Ibragimov, CRC Handbook of Lie Group Analysis of Differential Equations, V. 3, CRC Press, New York, 1996.

[8] B. D. Annin, V. O. Bytev, S. I. Senashov, Group properties of equations of elasticity and plasticity (Gruppovye svojstva uravnenij uprugosti i plastichnosti), Nauka, Novosibirsk, 1985.

[9] Michael M. Tung. FORM Matters: Fast Symbolic Computation under UNIX. arXiv:cs.SC/0409048 v1 27 Sep 2004

APPENDIX 1

The infinitesimal invariance criteria for PDE (1) is

$$\delta \frac{\partial n}{\partial t} + \delta u \frac{\partial n}{\partial x} + u\delta \frac{\partial n}{\partial x} + \delta v \frac{\partial n}{\partial y} + v\delta \frac{\partial n}{\partial y} + \delta w \frac{\partial n}{\partial z} + w\delta \frac{\partial n}{\partial z} - \tag{A1-1}$$

$$-a\delta u \frac{\partial n}{\partial u} - au\delta \frac{\partial n}{\partial u} - a\delta v \frac{\partial n}{\partial v} - av\delta \frac{\partial n}{\partial v} - a\delta w \frac{\partial n}{\partial w} - aw\delta \frac{\partial n}{\partial w} -$$

$$-3a\delta n - k(\delta \frac{\partial^2 n}{\partial u^2} + \delta \frac{\partial^2 n}{\partial v^2} + \delta \frac{\partial^2 n}{\partial w^2}) = 0;$$

Expressions for variations of derivatives are presented in APPENDIX 2.

We eliminate $\frac{\partial n}{\partial t}$ using original equation

$$\frac{\partial n}{\partial t} = -(u \frac{\partial n}{\partial x} + v \frac{\partial n}{\partial y} + w \frac{\partial n}{\partial z} - au \frac{\partial n}{\partial u} - av \frac{\partial n}{\partial v} - aw \frac{\partial n}{\partial w} - 3an - k(\frac{\partial^2 n}{\partial u^2} + \frac{\partial^2 n}{\partial v^2} + \frac{\partial^2 n}{\partial w^2})); \tag{A1-2}$$

Collecting similar terms, we obtain following equations:

$\frac{\partial n}{\partial x} \frac{\partial n}{\partial u}$

$$-2ku \frac{\partial^2}{\partial n \partial u} (\delta t) + 2k \frac{\partial^2}{\partial n \partial u} (\delta x) = 0; \tag{A1-3}$$

$\frac{\partial n}{\partial x} \frac{\partial n}{\partial u^2}$

$$-ku \frac{\partial^2}{\partial n^2} (\delta t) + k \frac{\partial^2}{\partial n^2} (\delta x) = 0; \tag{A1-4}$$

$\frac{\partial n}{\partial x} \frac{\partial n}{\partial v}$

$$-2ku \frac{\partial^2}{\partial n \partial v} (\delta t) + 2k \frac{\partial^2}{\partial n \partial v} (\delta x) = 0; \tag{A1-5}$$

$\frac{\partial n}{\partial x} \frac{\partial n}{\partial v^2}$

$$-ku \frac{\partial^2}{\partial n^2} (\delta t) + k \frac{\partial^2}{\partial n^2} (\delta x) = 0; \tag{A1-6}$$

$\frac{\partial n}{\partial x} \frac{\partial n}{\partial w}$

$$-2ku \frac{\partial^2}{\partial n \partial w} (\delta t) + 2k \frac{\partial^2}{\partial n \partial w} (\delta x) = 0; \tag{A1-7}$$

$\frac{\partial n}{\partial x} \frac{\partial n}{\partial w^2}$

$$-ku \frac{\partial^2}{\partial n^2} (\delta t) + k \frac{\partial^2}{\partial n^2} (\delta x) = 0; \tag{A1-8}$$

$\frac{\partial n}{\partial x}$

$$-auv \frac{\partial}{\partial v} (\delta t) - auw \frac{\partial}{\partial w} (\delta t) + 3aun \frac{\partial}{\partial n} (\delta t) + au \frac{\partial}{\partial u} (\delta x) - \tag{A1-9}$$

$$-au^2 \frac{\partial}{\partial u}(\delta t) + av \frac{\partial}{\partial v}(\delta x) + aw \frac{\partial}{\partial w}(\delta x) - 3an \frac{\partial}{\partial n}(\delta x) -$$

$$-ku \frac{\partial^2}{\partial u^2}(\delta t) - ku \frac{\partial^2}{\partial v^2}(\delta t) - ku \frac{\partial^2}{\partial w^2}(\delta t) + k \frac{\partial^2}{\partial u^2}(\delta x) +$$

$$+k \frac{\partial^2}{\partial v^2}(\delta x) + k \frac{\partial^2}{\partial w^2}(\delta x) + uv \frac{\partial}{\partial y}(\delta t) + uw \frac{\partial}{\partial z}(\delta t) -$$

$$-u \frac{\partial}{\partial x}(\delta x) + u \frac{\partial}{\partial t}(\delta t) + u^2 \frac{\partial}{\partial x}(\delta t) -$$

$$-v \frac{\partial}{\partial y}(\delta x) - w \frac{\partial}{\partial z}(\delta x) + \delta u - \frac{\partial}{\partial t}(\delta x) = 0;$$

$\dfrac{\partial n}{\partial y} \dfrac{\partial n}{\partial u}$

$$-2kv \frac{\partial^2}{\partial n \partial u}(\delta t) + 2k \frac{\partial^2}{\partial n \partial u}(\delta y) = 0; \qquad \text{(A1-10)}$$

$\dfrac{\partial n}{\partial y} \dfrac{\partial n}{\partial u^2}$

$$-kv \frac{\partial^2}{\partial n^2}(\delta t) + k \frac{\partial^2}{\partial n^2}(\delta y) = 0; \qquad \text{(A1-11)}$$

$\dfrac{\partial n}{\partial y} \dfrac{\partial n}{\partial v}$

$$-2kv \frac{\partial^2}{\partial n \partial v}(\delta t) + 2k \frac{\partial^2}{\partial n \partial v}(\delta y) = 0; \qquad \text{(A1-12)}$$

$\dfrac{\partial n}{\partial y} \dfrac{\partial n}{\partial v^2}$

$$-kv \frac{\partial^2}{\partial n^2}(\delta t) + k \frac{\partial^2}{\partial n^2}(\delta y) = 0; \qquad \text{(A1-13)}$$

$\dfrac{\partial n}{\partial y} \dfrac{\partial n}{\partial w}$

$$-2kv \frac{\partial^2}{\partial n \partial w}(\delta t) + 2k \frac{\partial^2}{\partial n \partial w}(\delta y) = 0; \qquad \text{(A1-14)}$$

$\dfrac{\partial n}{\partial y} \dfrac{\partial n}{\partial w^2}$

$$-kv \frac{\partial^2}{\partial n^2}(\delta t) + k \frac{\partial^2}{\partial n^2}(\delta y) = 0; \qquad \text{(A1-15)}$$

$\dfrac{\partial n}{\partial y}$

$$-auv \frac{\partial}{\partial u}(\delta t) + au \frac{\partial}{\partial u}(\delta y) - avw \frac{\partial}{\partial w}(\delta t) + 3avn \frac{\partial}{\partial n}(\delta t) + \qquad \text{(A1-16)}$$

$$+av \frac{\partial}{\partial v}(\delta y) - av^2 \frac{\partial}{\partial v}(\delta t) + aw \frac{\partial}{\partial w}(\delta y) - 3an \frac{\partial}{\partial n}(\delta y) -$$

$$-kv \frac{\partial^2}{\partial u^2} (\delta t) - kv \frac{\partial^2}{\partial v^2} (\delta t) - kv \frac{\partial^2}{\partial w^2} (\delta t) + k \frac{\partial^2}{\partial u^2} (\delta y) +$$

$$+ k \frac{\partial^2}{\partial v^2} (\delta y) + k \frac{\partial^2}{\partial w^2} (\delta y) + uv \frac{\partial}{\partial x} (\delta t) - u \frac{\partial}{\partial x} (\delta y) +$$

$$+ vw \frac{\partial}{\partial z} (\delta t) - v \frac{\partial}{\partial y} (\delta y) + v \frac{\partial}{\partial t} (\delta t) +$$

$$+ v^2 \frac{\partial}{\partial y} (\delta t) - w \frac{\partial}{\partial z} (\delta y) + \delta v - \frac{\partial}{\partial t} (\delta y) = 0;$$

$\dfrac{\partial n}{\partial z} \dfrac{\partial n}{\partial u}$

$$-2kw \frac{\partial^2}{\partial n \partial u} (\delta t) + 2k \frac{\partial^2}{\partial n \partial u} (\delta z) = 0; \tag{A1-17}$$

$\dfrac{\partial n}{\partial z} \dfrac{\partial n}{\partial u^2}$

$$-kw \frac{\partial^2}{\partial n^2} (\delta t) + k \frac{\partial^2}{\partial n^2} (\delta z) = 0; \tag{A1-18}$$

$\dfrac{\partial n}{\partial z} \dfrac{\partial n}{\partial v}$

$$-2kw \frac{\partial^2}{\partial n \partial v} (\delta t) + 2k \frac{\partial^2}{\partial n \partial v} (\delta z) = 0; \tag{A1-19}$$

$\dfrac{\partial n}{\partial z} \dfrac{\partial n}{\partial v^2}$

$$-kw \frac{\partial^2}{\partial n^2} (\delta t) + k \frac{\partial^2}{\partial n^2} (\delta z) = 0; \tag{A1-20}$$

$\dfrac{\partial n}{\partial z} \dfrac{\partial n}{\partial w}$

$$-2kw \frac{\partial^2}{\partial n \partial w} (\delta t) + 2k \frac{\partial^2}{\partial n \partial w} (\delta z) = 0; \tag{A1-21}$$

$\dfrac{\partial n}{\partial z} \dfrac{\partial n}{\partial w^2}$

$$-kw \frac{\partial^2}{\partial n^2} (\delta t) + k \frac{\partial^2}{\partial n^2} (\delta z) = 0; \tag{A1-22}$$

$\dfrac{\partial n}{\partial z}$

$$-auw \frac{\partial}{\partial u} (\delta t) + au \frac{\partial}{\partial u} (\delta z) - avw \frac{\partial}{\partial v} (\delta t) + av \frac{\partial}{\partial v} (\delta z) + \tag{A1-23}$$

$$+ 3awn \frac{\partial}{\partial n} (\delta t) + aw \frac{\partial}{\partial w} (\delta z) - aw^2 \frac{\partial}{\partial w} (\delta t) - 3an \frac{\partial}{\partial n} (\delta z) -$$

$$- kw \frac{\partial^2}{\partial v^2} (\delta t) - kw \frac{\partial^2}{\partial w^2} (\delta t) + k \frac{\partial^2}{\partial u^2} (\delta z) + k \frac{\partial^2}{\partial v^2} (\delta z) +$$

$$+ k \frac{\partial^2}{\partial w^2} (\delta z) + uw \frac{\partial}{\partial x} (\delta t) - u \frac{\partial}{\partial x} (\delta z) + vw \frac{\partial}{\partial y} (\delta t) - v \frac{\partial}{\partial y} (\delta z) -$$

$$- w \frac{\partial}{\partial z} (\delta z) + w \frac{\partial}{\partial t} (\delta t) + w^2 \frac{\partial}{\partial z} (\delta t) + \delta w - \frac{\partial}{\partial t} (\delta z) = 0;$$

$\dfrac{\partial n}{\partial u} \dfrac{\partial n}{\partial v}$

$$2aku \frac{\partial^2}{\partial n \partial v} (\delta t) + 2akv \frac{\partial^2}{\partial n \partial u} (\delta t) + 2k \frac{\partial^2}{\partial n \partial u} (\delta v) + 2k \frac{\partial^2}{\partial n \partial v} (\delta u) = 0; \tag{A1-24}$$

$\dfrac{\partial n}{\partial u} \dfrac{\partial n}{\partial v^2}$

$$aku \frac{\partial^2}{\partial n^2} (\delta t) + k \frac{\partial^2}{\partial n^2} (\delta u) = 0; \tag{A1-25}$$

$\dfrac{\partial n}{\partial u} \dfrac{\partial n}{\partial w}$

$$2aku \frac{\partial^2}{\partial n \partial w} (\delta t) + 2akw \frac{\partial^2}{\partial n \partial u} (\delta t) + 2k \frac{\partial^2}{\partial n \partial u} (\delta w) + 2k \frac{\partial^2}{\partial n \partial w} (\delta u) = 0; \tag{A1-26}$$

$\dfrac{\partial n}{\partial u} \dfrac{\partial n}{\partial w^2}$

$$aku \frac{\partial^2}{\partial n^2} (\delta t) + k \frac{\partial^2}{\partial n^2} (\delta u) = 0; \tag{A1-27}$$

$\dfrac{\partial n}{\partial u} \dfrac{\partial^2 n}{\partial u^2}$

$$2k \frac{\partial}{\partial n} (\delta u) + 2k^2 \frac{\partial^2}{\partial n \partial u} (\delta t) = 0; \tag{A1-28}$$

$\dfrac{\partial n}{\partial u} \dfrac{\partial^2 n}{\partial v^2}$

$$2k^2 \frac{\partial^2}{\partial n \partial u} (\delta t) = 0; \tag{A1-29}$$

$\dfrac{\partial n}{\partial u} \dfrac{\partial^2 n}{\partial w^2}$

$$2k^2 \frac{\partial^2}{\partial n \partial u} (\delta t) = 0; \tag{A1-30}$$

$\dfrac{\partial n}{\partial u} \dfrac{\partial^2 n}{\partial u \partial v}$

$$2k \frac{\partial}{\partial n} (\delta v) = 0; \tag{A1-31}$$

$\dfrac{\partial n}{\partial u} \dfrac{\partial^2 n}{\partial u \partial w}$

$$2k \frac{\partial}{\partial n} (\delta w) = 0; \tag{A1-32}$$

$\dfrac{\partial n}{\partial u} \dfrac{\partial^2 n}{\partial u \partial x}$

$$2k \frac{\partial}{\partial n} (\delta x) = 0; \tag{A1-33}$$

$\dfrac{\partial n}{\partial u} \quad \dfrac{\partial^2 n}{\partial u \partial y}$

$$2k \frac{\partial}{\partial n} (\delta y) = 0; \tag{A1-34}$$

$\dfrac{\partial n}{\partial u} \quad \dfrac{\partial^2 n}{\partial u \partial z}$

$$k \frac{\partial}{\partial n} (\delta z) = 0; \tag{A1-35}$$

$\dfrac{\partial n}{\partial u} \quad \dfrac{\partial^2 n}{\partial t \partial u}$

$$2k \frac{\partial}{\partial n} (\delta t) = 0; \tag{A1-36}$$

$\dfrac{\partial n}{\partial u}$

$$aku \frac{\partial^2}{\partial u^2} (\delta t) + aku \frac{\partial^2}{\partial v^2} (\delta t) + aku \frac{\partial^2}{\partial w^2} (\delta t) + 6akn \frac{\partial^2}{\partial n \partial u} (\delta t) - \tag{A1-37}$$

$$- auv \frac{\partial}{\partial y} (\delta t) - auw \frac{\partial}{\partial z} (\delta t) + au \frac{\partial}{\partial u} (\delta u) - au \frac{\partial}{\partial t} (\delta t) -$$

$$- au^2 \frac{\partial}{\partial x} (\delta t) + av \frac{\partial}{\partial v} (\delta u) + aw \frac{\partial}{\partial w} (\delta u) - 3an \frac{\partial}{\partial n} (\delta u) -$$

$$- a \delta u + a^2 uv \frac{\partial}{\partial v} (\delta t) + a^2 uw \frac{\partial}{\partial w} (\delta t) - 3a^2 un \frac{\partial}{\partial n} (\delta t) + a^2 u^2 \frac{\partial}{\partial u} (\delta t) +$$

$$+ k \frac{\partial^2 n}{\partial u \partial z} \frac{\partial}{\partial n} (\delta z) - 2k \frac{\partial^2}{\partial n \partial u} (\delta n) +$$

$$+ k \frac{\partial^2}{\partial u^2} (\delta u) + k \frac{\partial^2}{\partial v^2} (\delta u) + k \frac{\partial^2}{\partial w^2} (\delta u) -$$

$$- u \frac{\partial}{\partial x} (\delta u) - v \frac{\partial}{\partial y} (\delta u) - w \frac{\partial}{\partial z} (\delta u) - \frac{\partial}{\partial t} (\delta u) = 0;$$

$\dfrac{\partial n}{\partial u^2} \quad \dfrac{\partial n}{\partial v}$

$$akv \frac{\partial^2}{\partial n^2} (\delta t) + k \frac{\partial^2}{\partial n^2} (\delta v) = 0; \tag{A1-38}$$

$\dfrac{\partial n}{\partial u^2} \quad \dfrac{\partial n}{\partial w}$

$$akw \frac{\partial^2}{\partial n^2} (\delta t) + k \frac{\partial^2}{\partial n^2} (\delta w) = 0; \tag{A1-39}$$

$\dfrac{\partial n}{\partial u^2} \quad \dfrac{\partial^2 n}{\partial u^2}$

$$k^2 \frac{\partial^2}{\partial n^2} (\delta t) = 0; \tag{A1-40}$$

$$\frac{\partial n}{\partial u^2}\ \frac{\partial^2 n}{\partial v^2}$$

$$k^2\ \frac{\partial^2}{\partial n^2}\ (\delta t) = 0; \qquad\qquad (A1\text{-}41)$$

$$\frac{\partial n}{\partial u^2}\ \frac{\partial^2 n}{\partial w^2}$$

$$k^2\ \frac{\partial^2}{\partial n^2}\ (\delta t) = 0; \qquad\qquad (A1\text{-}42)$$

$$\frac{\partial n}{\partial u^2}$$

$$2aku\ \frac{\partial^2}{\partial n\partial u}\ (\delta t) + 3akn\ \frac{\partial^2}{\partial n^2}\ (\delta t) - k\ \frac{\partial^2}{\partial n^2}\ (\delta n) + 2k\ \frac{\partial^2}{\partial n\partial u}\ (\delta u) = 0; \qquad\qquad (A1\text{-}43)$$

$$\frac{\partial n}{\partial u^3}$$

$$aku\ \frac{\partial^2}{\partial n^2}\ (\delta t) + k\ \frac{\partial^2}{\partial n^2}\ (\delta u) = 0; \qquad\qquad (A1\text{-}44)$$

$$\frac{\partial n}{\partial v}\ \frac{\partial n}{\partial w}$$

$$2akv\ \frac{\partial^2}{\partial n\partial w}\ (\delta t) + 2akw\ \frac{\partial^2}{\partial n\partial v}\ (\delta t) + 2k\ \frac{\partial^2}{\partial n\partial v}\ (\delta w) + 2k\ \frac{\partial^2}{\partial n\partial w}\ (\delta v) = 0; \qquad\qquad (A1\text{-}45)$$

$$\frac{\partial n}{\partial v}\ \frac{\partial n}{\partial w^2}$$

$$akv\ \frac{\partial^2}{\partial n^2}\ (\delta t) + k\ \frac{\partial^2}{\partial n^2}\ (\delta v) = 0; \qquad\qquad (A1\text{-}46)$$

$$\frac{\partial n}{\partial v}\ \frac{\partial^2 n}{\partial u^2}$$

$$2k^2\ \frac{\partial^2}{\partial n\partial v}\ (\delta t) = 0; \qquad\qquad (A1\text{-}47)$$

$$\frac{\partial n}{\partial v}\ \frac{\partial^2 n}{\partial v^2}$$

$$2k\ \frac{\partial}{\partial n}\ (\delta v) + 2k^2\ \frac{\partial^2}{\partial n\partial v}\ (\delta t) = 0; \qquad\qquad (A1\text{-}48)$$

$$\frac{\partial n}{\partial v}\ \frac{\partial^2 n}{\partial w^2}$$

$$2k^2\ \frac{\partial^2}{\partial n\partial v}\ (\delta t) = 0; \qquad\qquad (A1\text{-}49)$$

$$\frac{\partial n}{\partial v}\ \frac{\partial^2 n}{\partial u\partial v}$$

$$2k\ \frac{\partial}{\partial n}\ (\delta u) = 0; \qquad\qquad (A1\text{-}50)$$

$$\frac{\partial n}{\partial v}\ \frac{\partial^2 n}{\partial v\partial w}$$

$$2k \frac{\partial}{\partial n} (\delta w) = 0; \tag{A1-51}$$

$\dfrac{\partial n}{\partial v} \quad \dfrac{\partial^2 n}{\partial v \partial x}$

$$2k \frac{\partial}{\partial n} (\delta x) = 0; \tag{A1-52}$$

$\dfrac{\partial n}{\partial v} \quad \dfrac{\partial^2 n}{\partial v \partial y}$

$$2k \frac{\partial}{\partial n} (\delta y) = 0; \tag{A1-53}$$

$\dfrac{\partial n}{\partial v} \quad \dfrac{\partial^2 n}{\partial v \partial z}$

$$k \frac{\partial}{\partial n} (\delta z) = 0; \tag{A1-54}$$

$\dfrac{\partial n}{\partial v} \quad \dfrac{\partial^2 n}{\partial t \partial v}$

$$2k \frac{\partial}{\partial n} (\delta t) = 0; \tag{A1-55}$$

$\dfrac{\partial n}{\partial v}$

$$akv \frac{\partial^2}{\partial u^2} (\delta t) + akv \frac{\partial^2}{\partial v^2} (\delta t) + akv \frac{\partial^2}{\partial w^2} (\delta t) + \tag{A1-56}$$

$$+ 6akn \frac{\partial^2}{\partial n \partial v} (\delta t) - auv \frac{\partial}{\partial x} (\delta t) + au \frac{\partial}{\partial u} (\delta v) - avw \frac{\partial}{\partial z} (\delta t) +$$

$$+ av \frac{\partial}{\partial v} (\delta v) - av \frac{\partial}{\partial t} (\delta t) - av^2 \frac{\partial}{\partial y} (\delta t) + aw \frac{\partial}{\partial w} (\delta v) -$$

$$- 3an \frac{\partial}{\partial n} (\delta v) - a\delta v + a^2 uv \frac{\partial}{\partial u} (\delta t) + a^2 vw \frac{\partial}{\partial w} (\delta t) - 3a^2 vn \frac{\partial}{\partial n} (\delta t) +$$

$$+ a^2 v^2 \frac{\partial}{\partial v} (\delta t) + k \frac{\partial^2 n}{\partial v \partial z} \frac{\partial}{\partial n} (\delta z) + k \frac{\partial^2}{\partial u^2} (\delta v) - 2k \frac{\partial^2}{\partial n \partial v} (\delta n) +$$

$$+ k \frac{\partial^2}{\partial v^2} (\delta v) + k \frac{\partial^2}{\partial w^2} (\delta v) - u \frac{\partial}{\partial x} (\delta v) - v \frac{\partial}{\partial y} (\delta v) - w \frac{\partial}{\partial z} (\delta v) - \frac{\partial}{\partial t} (\delta v) = 0;$$

$\dfrac{\partial n}{\partial v^2} \quad \dfrac{\partial n}{\partial w}$

$$akw \frac{\partial^2}{\partial n^2} (\delta t) + k \frac{\partial^2}{\partial n^2} (\delta w) = 0; \tag{A1-57}$$

$\dfrac{\partial n}{\partial v^2} \quad \dfrac{\partial^2 n}{\partial u^2}$

$$k^2 \frac{\partial^2}{\partial n^2} (\delta t) = 0; \tag{A1-58}$$

$\dfrac{\partial n}{\partial v^2} \quad \dfrac{\partial^2 n}{\partial v^2}$

$$k^2 \frac{\partial^2}{\partial n^2} (\delta t) = 0; \tag{A1-59}$$

$\dfrac{\partial n}{\partial v^2} \dfrac{\partial^2 n}{\partial w^2}$

$$k^2 \frac{\partial^2}{\partial n^2} (\delta t) = 0; \tag{A1-60}$$

$\dfrac{\partial n}{\partial v^2}$

$$2akv \frac{\partial^2}{\partial n \partial v} (\delta t) + 3akn \frac{\partial^2}{\partial n^2} (\delta t) - k \frac{\partial^2}{\partial n^2} (\delta n) + 2k \frac{\partial^2}{\partial n \partial v} (\delta v) = 0; \tag{A1-61}$$

$\dfrac{\partial n}{\partial v^3}$

$$akv \frac{\partial^2}{\partial n^2} (\delta t) + k \frac{\partial^2}{\partial n^2} (\delta v) = 0; \tag{A1-62}$$

$\dfrac{\partial n}{\partial w} \dfrac{\partial^2 n}{\partial u^2}$

$$2k^2 \frac{\partial^2}{\partial n \partial w} (\delta t) = 0; \tag{A1-63}$$

$\dfrac{\partial n}{\partial w} \dfrac{\partial^2 n}{\partial v^2}$

$$2k^2 \frac{\partial^2}{\partial n \partial w} (\delta t) = 0; \tag{A1-64}$$

$\dfrac{\partial n}{\partial w} \dfrac{\partial^2 n}{\partial w^2}$

$$2k \frac{\partial}{\partial n} (\delta w) + 2k^2 \frac{\partial^2}{\partial n \partial w} (\delta t) = 0; \tag{A1-65}$$

$\dfrac{\partial n}{\partial w} \dfrac{\partial^2 n}{\partial u \partial w}$

$$2k \frac{\partial}{\partial n} (\delta u) = 0; \tag{A1-66}$$

$\dfrac{\partial n}{\partial w} \dfrac{\partial^2 n}{\partial v \partial w}$

$$2k \frac{\partial}{\partial n} (\delta v) = 0; \tag{A1-67}$$

$\dfrac{\partial n}{\partial w} \dfrac{\partial^2 n}{\partial w \partial x}$

$$2k \frac{\partial}{\partial n} (\delta x) = 0; \tag{A1-68}$$

$\dfrac{\partial n}{\partial w} \dfrac{\partial^2 n}{\partial w \partial y}$

$$2k \frac{\partial}{\partial n} (\delta y) = 0; \tag{A1-69}$$

$\dfrac{\partial n}{\partial w}\dfrac{\partial^2 n}{\partial w \partial z}$

$$k\,\frac{\partial}{\partial n}\,(\delta z) = 0; \qquad\qquad (A1\text{-}70)$$

$\dfrac{\partial n}{\partial w}\dfrac{\partial^2 n}{\partial t \partial w}$

$$2k\,\frac{\partial}{\partial n}\,(\delta t) = 0; \qquad\qquad (A1\text{-}71)$$

$\dfrac{\partial n}{\partial w}$

$$akw\,\frac{\partial^2}{\partial u^2}\,(\delta t) + akw\,\frac{\partial^2}{\partial v^2}\,(\delta t) + akw\,\frac{\partial^2}{\partial w^2}\,(\delta t) + 6akn\,\frac{\partial^2}{\partial n \partial w}\,(\delta t) - \qquad (A1\text{-}72)$$

$$-auw\,\frac{\partial}{\partial x}\,(\delta t) + au\,\frac{\partial}{\partial u}\,(\delta w) - avw\,\frac{\partial}{\partial y}\,(\delta t) + av\,\frac{\partial}{\partial v}\,(\delta w) + aw\,\frac{\partial}{\partial w}\,(\delta w) - aw\,\frac{\partial}{\partial t}\,(\delta t) - $$

$$-aw^2\,\frac{\partial}{\partial z}\,(\delta t) - 3an\,\frac{\partial}{\partial n}\,(\delta w) - a\delta w + a^2 uw\,\frac{\partial}{\partial u}\,(\delta t) + a^2 vw\,\frac{\partial}{\partial v}\,(\delta t) - $$

$$-3a^2 wn\,\frac{\partial}{\partial n}\,(\delta t) + a^2 w^2\,\frac{\partial}{\partial w}\,(\delta t) + k\,\frac{\partial^2 n}{\partial w \partial z}\,\frac{\partial}{\partial n}\,(\delta z) + k\,\frac{\partial^2}{\partial u^2}\,(\delta w) + k\,\frac{\partial^2}{\partial v^2}\,(\delta w) - $$

$$-2k\,\frac{\partial^2}{\partial n \partial w}\,(\delta n) + k\,\frac{\partial^2}{\partial w^2}\,(\delta w) - u\,\frac{\partial}{\partial x}\,(\delta w) - v\,\frac{\partial}{\partial y}\,(\delta w) - w\,\frac{\partial}{\partial z}\,(\delta w) - \frac{\partial}{\partial t}\,(\delta w) = 0; $$

$\dfrac{\partial n}{\partial w^2}\dfrac{\partial^2 n}{\partial u^2}$

$$k^2\,\frac{\partial^2}{\partial n^2}\,(\delta t) = 0; \qquad\qquad (A1\text{-}73)$$

$\dfrac{\partial n}{\partial w^2}\dfrac{\partial^2 n}{\partial v^2}$

$$k^2\,\frac{\partial^2}{\partial n^2}\,(\delta t) = 0; \qquad\qquad (A1\text{-}74)$$

$\dfrac{\partial n}{\partial w^2}\dfrac{\partial^2 n}{\partial w^2}$

$$k^2\,\frac{\partial^2}{\partial n^2}\,(\delta t) = 0; \qquad\qquad (A1\text{-}75)$$

$\dfrac{\partial n}{\partial w^2}$

$$2akw\,\frac{\partial^2}{\partial n \partial w}\,(\delta t) + 3akn\,\frac{\partial^2}{\partial n^2}\,(\delta t) - k\,\frac{\partial^2}{\partial n^2}\,(\delta n) + 2k\,\frac{\partial^2}{\partial n \partial w}\,(\delta w) = 0; \qquad (A1\text{-}76)$$

$\dfrac{\partial n}{\partial w^3}$

$$akw\,\frac{\partial^2}{\partial n^2}\,(\delta t) + k\,\frac{\partial^2}{\partial n^2}\,(\delta w) = 0; \qquad\qquad (A1\text{-}77)$$

$\dfrac{\partial^2 n}{\partial u^2}$

$$aku\frac{\partial}{\partial u}(\delta t) + akv\frac{\partial}{\partial v}(\delta t) + akw\frac{\partial}{\partial w}(\delta t) - 3akn\frac{\partial}{\partial n}(\delta t) - \tag{A1-78}$$

$$-ku\frac{\partial}{\partial x}(\delta t) - kv\frac{\partial}{\partial y}(\delta t) - kw\frac{\partial}{\partial z}(\delta t) + 2k\frac{\partial}{\partial u}(\delta u) - k\frac{\partial}{\partial t}(\delta t) +$$

$$+k^2\frac{\partial^2}{\partial u^2}(\delta t) + k^2\frac{\partial^2}{\partial v^2}(\delta t) + k^2\frac{\partial^2}{\partial w^2}(\delta t) = 0;$$

$\dfrac{\partial^2 n}{\partial v^2}$

$$aku\frac{\partial}{\partial u}(\delta t) + akv\frac{\partial}{\partial v}(\delta t) + akw\frac{\partial}{\partial w}(\delta t) - 3akn\frac{\partial}{\partial n}(\delta t) - \tag{A1-79}$$

$$-ku\frac{\partial}{\partial x}(\delta t) - kv\frac{\partial}{\partial y}(\delta t) - kw\frac{\partial}{\partial z}(\delta t) + 2k\frac{\partial}{\partial v}(\delta v) - k\frac{\partial}{\partial t}(\delta t) +$$

$$+k^2\frac{\partial^2}{\partial u^2}(\delta t) + k^2\frac{\partial^2}{\partial v^2}(\delta t) + k^2\frac{\partial^2}{\partial w^2}(\delta t) = 0;$$

$\dfrac{\partial^2 n}{\partial w^2}$

$$aku\frac{\partial}{\partial u}(\delta t) + akv\frac{\partial}{\partial v}(\delta t) + akw\frac{\partial}{\partial w}(\delta t) - 3akn\frac{\partial}{\partial n}(\delta t) - \tag{A1-80}$$

$$-ku\frac{\partial}{\partial x}(\delta t) - kv\frac{\partial}{\partial y}(\delta t) - kw\frac{\partial}{\partial z}(\delta t) + 2k\frac{\partial}{\partial w}(\delta w) - k\frac{\partial}{\partial t}(\delta t) +$$

$$+k^2\frac{\partial^2}{\partial u^2}(\delta t) + k^2\frac{\partial^2}{\partial v^2}(\delta t) + k^2\frac{\partial^2}{\partial w^2}(\delta t) = 0;$$

$\dfrac{\partial^2 n}{\partial u\partial v}$

$$2k\frac{\partial}{\partial v}(\delta u) + 2k\frac{\partial}{\partial u}(\delta v) = 0; \tag{A1-81}$$

$\dfrac{\partial^2 n}{\partial u\partial w}$

$$2k\frac{\partial}{\partial w}(\delta u) + 2k\frac{\partial}{\partial u}(\delta w) = 0; \tag{A1-82}$$

$\dfrac{\partial^2 n}{\partial v\partial w}$

$$2k\frac{\partial}{\partial w}(\delta v) + 2k\frac{\partial}{\partial v}(\delta w) = 0; \tag{A1-83}$$

$\dfrac{\partial^2 n}{\partial u\partial x}$

$$2k\frac{\partial}{\partial u}(\delta x) = 0; \tag{A1-84}$$

$\dfrac{\partial^2 n}{\partial u\partial y}$

$$2k\frac{\partial}{\partial u}(\delta y) = 0; \tag{A1-85}$$

$\dfrac{\partial^2 n}{\partial u \partial z}$

$$k \frac{\partial}{\partial u} (\delta z) = 0; \qquad\qquad \text{(A1-86)}$$

$\dfrac{\partial^2 n}{\partial v \partial x}$

$$2k \frac{\partial}{\partial v} (\delta x) = 0; \qquad\qquad \text{(A1-87)}$$

$\dfrac{\partial^2 n}{\partial v \partial y}$

$$2k \frac{\partial}{\partial v} (\delta y) = 0; \qquad\qquad \text{(A1-88)}$$

$\dfrac{\partial^2 n}{\partial v \partial z}$

$$k \frac{\partial}{\partial v} (\delta z) = 0; \qquad\qquad \text{(A1-89)}$$

$\dfrac{\partial^2 n}{\partial w \partial x}$

$$2k \frac{\partial}{\partial w} (\delta x) = 0; \qquad\qquad \text{(A1-90)}$$

$\dfrac{\partial^2 n}{\partial w \partial y}$

$$2k \frac{\partial}{\partial w} (\delta y) = 0; \qquad\qquad \text{(A1-91)}$$

$\dfrac{\partial^2 n}{\partial w \partial z}$

$$k \frac{\partial}{\partial w} (\delta z) = 0; \qquad\qquad \text{(A1-92)}$$

$\dfrac{\partial^2 n}{\partial t \partial u}$

$$2k \frac{\partial}{\partial u} (\delta t) = 0; \qquad\qquad \text{(A1-93)}$$

$\dfrac{\partial^2 n}{\partial t \partial v}$

$$2k \frac{\partial}{\partial v} (\delta t) = 0; \qquad\qquad \text{(A1-94)}$$

$\dfrac{\partial^2 n}{\partial t \partial w}$

$$2k \frac{\partial}{\partial w} (\delta t) = 0; \qquad\qquad \text{(A1-95)}$$

$$3akn \frac{\partial^2}{\partial u^2} (\delta t) + 3akn \frac{\partial^2}{\partial v^2} (\delta t) + 3akn \frac{\partial^2}{\partial w^2} (\delta t) - \qquad \text{(A1-96)}$$

$$-3aun\frac{\partial}{\partial x}(\delta t)-au\frac{\partial}{\partial u}(\delta n)-3avn\frac{\partial}{\partial y}(\delta t)-av\frac{\partial}{\partial v}(\delta n)-3awn\frac{\partial}{\partial z}(\delta t)-aw\frac{\partial}{\partial w}(\delta n)+$$

$$+3an\frac{\partial}{\partial n}(\delta n)-3an\frac{\partial}{\partial t}(\delta t)-3a\delta n+3a^2un\frac{\partial}{\partial u}(\delta t)+3a^2vn\frac{\partial}{\partial v}(\delta t)+3a^2wn\frac{\partial}{\partial w}(\delta t)-9a^2n^2\frac{\partial}{\partial n}(\delta t)+$$

$$+k\frac{\partial^2 n}{\partial u\partial z}\frac{\partial}{\partial u}(\delta z)+k\frac{\partial^2 n}{\partial v\partial z}\frac{\partial}{\partial v}(\delta z)+k\frac{\partial^2 n}{\partial w\partial z}\frac{\partial}{\partial w}(\delta z)-$$

$$-k\frac{\partial^2}{\partial u^2}(\delta n)-k\frac{\partial^2}{\partial v^2}(\delta n)-k\frac{\partial^2}{\partial w^2}(\delta n)+$$

$$+u\frac{\partial}{\partial x}(\delta n)+v\frac{\partial}{\partial y}(\delta n)+w\frac{\partial}{\partial z}(\delta n)+\frac{\partial}{\partial t}(\delta n)=0.$$

From (A1-31 - A1-36), (A1-50 - A1-55), (A1-66 - A1-71), (A1-84 - A1-95) we see, that $\delta x=\delta x(x,y,z,t)$; $\delta y=\delta y(x,y,z,t)$; $\delta t=\delta t(x,y,z,t)$. We use these expressions to simplify the rest of equations (A1-3 - A1-96).

$$uv\frac{\partial}{\partial y}(\delta t)+uw\frac{\partial}{\partial z}(\delta t)-u\frac{\partial}{\partial x}(\delta x)+u\frac{\partial}{\partial t}(\delta t)+ \tag{A1-97}$$

$$+u^2\frac{\partial}{\partial x}(\delta t)-v\frac{\partial}{\partial y}(\delta x)-w\frac{\partial}{\partial z}(\delta x)+\delta u-\frac{\partial}{\partial t}(\delta x)=0;$$

$$uv\frac{\partial}{\partial x}(\delta t)-u\frac{\partial}{\partial x}(\delta y)+vw\frac{\partial}{\partial z}(\delta t)-v\frac{\partial}{\partial y}(\delta y)+ \tag{A1-98}$$

$$+v\frac{\partial}{\partial t}(\delta t)+v^2\frac{\partial}{\partial y}(\delta t)-w\frac{\partial}{\partial z}(\delta y)+\delta v-\frac{\partial}{\partial t}(\delta y)=0;$$

$$uw\frac{\partial}{\partial x}(\delta t)-u\frac{\partial}{\partial x}(\delta z)+vw\frac{\partial}{\partial y}(\delta t)-v\frac{\partial}{\partial y}(\delta z)- \tag{A1-99}$$

$$-w\frac{\partial}{\partial z}(\delta z)+w\frac{\partial}{\partial t}(\delta t)+w^2\frac{\partial}{\partial z}(\delta t)+\delta w-\frac{\partial}{\partial t}(\delta z)=0;$$

$$-auv\frac{\partial}{\partial y}(\delta t)-auw\frac{\partial}{\partial z}(\delta t)+au\frac{\partial}{\partial u}(\delta u)-au\frac{\partial}{\partial t}(\delta t)-au^2\frac{\partial}{\partial x}(\delta t)+ \tag{A1-100}$$

$$+av\frac{\partial}{\partial v}(\delta u)+aw\frac{\partial}{\partial w}(\delta u)-a\delta u-2k\frac{\partial^2}{\partial n\partial u}(\delta n)+$$

$$+k\frac{\partial^2}{\partial u^2}(\delta u)+k\frac{\partial^2}{\partial v^2}(\delta u)+k\frac{\partial^2}{\partial w^2}(\delta u)-u\frac{\partial}{\partial x}(\delta u)-v\frac{\partial}{\partial y}(\delta u)-w\frac{\partial}{\partial z}(\delta u)-\frac{\partial}{\partial t}(\delta u)=0;$$

$$-k\frac{\partial^2}{\partial n^2}(\delta n)=0; \tag{A1-101}$$

$$-auv\frac{\partial}{\partial x}(\delta t)+au\frac{\partial}{\partial u}(\delta v)-avw\frac{\partial}{\partial z}(\delta t)+av\frac{\partial}{\partial v}(\delta v)-av\frac{\partial}{\partial t}(\delta t)- \tag{A1-102}$$

$$-av^2\frac{\partial}{\partial y}(\delta t)+aw\frac{\partial}{\partial w}(\delta v)-a\delta v+k\frac{\partial^2}{\partial u^2}(\delta v)-2k\frac{\partial^2}{\partial n\partial v}(\delta n)+$$

$$+k\frac{\partial^2}{\partial v^2}(\delta v)+k\frac{\partial^2}{\partial w^2}(\delta v)-u\frac{\partial}{\partial x}(\delta v)-v\frac{\partial}{\partial y}(\delta v)-w\frac{\partial}{\partial z}(\delta v)-\frac{\partial}{\partial t}(\delta v)=0;$$

$$-k\frac{\partial^2}{\partial n^2}(\delta n)=0; \qquad (\text{A1-103})$$

$$-auw\frac{\partial}{\partial x}(\delta t)+au\frac{\partial}{\partial u}(\delta w)-avw\frac{\partial}{\partial y}(\delta t)+av\frac{\partial}{\partial v}(\delta w)+aw\frac{\partial}{\partial w}(\delta w)- \qquad (\text{A1-104})$$

$$-aw\frac{\partial}{\partial t}(\delta t)-aw^2\frac{\partial}{\partial z}(\delta t)-a\delta w+k\frac{\partial^2}{\partial u^2}(\delta w)+k\frac{\partial^2}{\partial v^2}(\delta w)-2k\frac{\partial^2}{\partial n\partial w}(\delta n)+$$

$$+k\frac{\partial^2}{\partial w^2}(\delta w)-u\frac{\partial}{\partial x}(\delta w)-v\frac{\partial}{\partial y}(\delta w)-w\frac{\partial}{\partial z}(\delta w)-\frac{\partial}{\partial t}(\delta w)=0;$$

$$-k\frac{\partial^2}{\partial n^2}(\delta n)=0; \qquad (\text{A1-105})$$

$$-ku\frac{\partial}{\partial x}(\delta t)-kv\frac{\partial}{\partial y}(\delta t)-kw\frac{\partial}{\partial z}(\delta t)+2k\frac{\partial}{\partial u}(\delta u)-k\frac{\partial}{\partial t}(\delta t)=0; \qquad (\text{A1-106})$$

$$-ku\frac{\partial}{\partial x}(\delta t)-kv\frac{\partial}{\partial y}(\delta t)-kw\frac{\partial}{\partial z}(\delta t)+2k\frac{\partial}{\partial v}(\delta v)-k\frac{\partial}{\partial t}(\delta t)=0; \qquad (\text{A1-107})$$

$$-ku\frac{\partial}{\partial x}(\delta t)-kv\frac{\partial}{\partial y}(\delta t)-kw\frac{\partial}{\partial z}(\delta t)+2k\frac{\partial}{\partial w}(\delta w)-k\frac{\partial}{\partial t}(\delta t)=0; \qquad (\text{A1-108})$$

$$2k\frac{\partial}{\partial v}(\delta u)+2k\frac{\partial}{\partial u}(\delta v)=0; \qquad (\text{A1-109})$$

$$2k\frac{\partial}{\partial w}(\delta u)+2k\frac{\partial}{\partial u}(\delta w)=0; \qquad (\text{A1-110})$$

$$2k\frac{\partial}{\partial w}(\delta v)+2k\frac{\partial}{\partial v}(\delta w)=0; \qquad (\text{A1-111})$$

$$-3aun\frac{\partial}{\partial x}(\delta t)-au\frac{\partial}{\partial u}(\delta n)-3avn\frac{\partial}{\partial y}(\delta t)-av\frac{\partial}{\partial v}(\delta n)-3awn\frac{\partial}{\partial z}(\delta t)- \qquad (\text{A1-112})$$

$$-aw\frac{\partial}{\partial w}(\delta n)+3an\frac{\partial}{\partial n}(\delta n)-3an\frac{\partial}{\partial t}(\delta t)-3a\delta n-k\frac{\partial^2}{\partial u^2}(\delta n)-$$

$$-k\frac{\partial^2}{\partial v^2}(\delta n)-k\frac{\partial^2}{\partial w^2}(\delta n)+u\frac{\partial}{\partial x}(\delta n)+v\frac{\partial}{\partial y}(\delta n)+w\frac{\partial}{\partial z}(\delta n)+\frac{\partial}{\partial t}(\delta n)=0.$$

From (A1-101), (A1-103) and (A1-105) we conclude, that

$$\delta n=A+nB; \qquad (\text{A1-113})$$

where $A=A(x,y,z,u,v,w,t)$, $B=B(x,y,z,u,v,w,t)$. Using these expressions, we simplify the rest of equations (A1-97 - A1-112).

$$uv\frac{\partial}{\partial y}(\delta t)+uw\frac{\partial}{\partial z}(\delta t)-u\frac{\partial}{\partial x}(\delta x)+u\frac{\partial}{\partial t}(\delta t)+ \qquad (\text{A1-114})$$

$$+u^2\frac{\partial}{\partial x}(\delta t)-v\frac{\partial}{\partial y}(\delta x)-w\frac{\partial}{\partial z}(\delta x)+\delta u-\frac{\partial}{\partial t}(\delta x)=0;$$

$$uv \frac{\partial}{\partial x}(\delta t) - u \frac{\partial}{\partial x}(\delta y) + vw \frac{\partial}{\partial z}(\delta t) - v \frac{\partial}{\partial y}(\delta y) + \tag{A1-115}$$

$$+ v \frac{\partial}{\partial t}(\delta t) + v^2 \frac{\partial}{\partial y}(\delta t) - w \frac{\partial}{\partial z}(\delta y) + \delta v - \frac{\partial}{\partial t}(\delta y) = 0;$$

$$uw \frac{\partial}{\partial x}(\delta t) - u \frac{\partial}{\partial x}(\delta z) + vw \frac{\partial}{\partial y}(\delta t) - v \frac{\partial}{\partial y}(\delta z) - \tag{A1-116}$$

$$- w \frac{\partial}{\partial z}(\delta z) + w \frac{\partial}{\partial t}(\delta t) + w^2 \frac{\partial}{\partial z}(\delta t) + \delta w - \frac{\partial}{\partial t}(\delta z) = 0;$$

$$-auv \frac{\partial}{\partial y}(\delta t) - auw \frac{\partial}{\partial z}(\delta t) + au \frac{\partial}{\partial u}(\delta u) - au \frac{\partial}{\partial t}(\delta t) - au^2 \frac{\partial}{\partial x}(\delta t) + \tag{A1-117}$$

$$+ av \frac{\partial}{\partial v}(\delta u) + aw \frac{\partial}{\partial w}(\delta u) - a\delta u - 2k \frac{\partial B}{\partial u} +$$

$$+ k \frac{\partial^2}{\partial u^2}(\delta u) + k \frac{\partial^2}{\partial v^2}(\delta u) + k \frac{\partial^2}{\partial w^2}(\delta u) -$$

$$- u \frac{\partial}{\partial x}(\delta u) - v \frac{\partial}{\partial y}(\delta u) - w \frac{\partial}{\partial z}(\delta u) - \frac{\partial}{\partial t}(\delta u) = 0;$$

$$-auv \frac{\partial}{\partial x}(\delta t) + au \frac{\partial}{\partial u}(\delta v) - avw \frac{\partial}{\partial z}(\delta t) + av \frac{\partial}{\partial v}(\delta v) - av \frac{\partial}{\partial t}(\delta t) - \tag{A1-118}$$

$$- av^2 \frac{\partial}{\partial y}(\delta t) + aw \frac{\partial}{\partial w}(\delta v) - a\delta v - 2k \frac{\partial B}{\partial v} +$$

$$+ k \frac{\partial^2}{\partial u^2}(\delta v) + k \frac{\partial^2}{\partial v^2}(\delta v) + k \frac{\partial^2}{\partial w^2}(\delta v) -$$

$$- u \frac{\partial}{\partial x}(\delta v) - v \frac{\partial}{\partial y}(\delta v) - w \frac{\partial}{\partial z}(\delta v) - \frac{\partial}{\partial t}(\delta v) = 0;$$

$$-auw \frac{\partial}{\partial x}(\delta t) + au \frac{\partial}{\partial u}(\delta w) - avw \frac{\partial}{\partial y}(\delta t) + av \frac{\partial}{\partial v}(\delta w) + aw \frac{\partial}{\partial w}(\delta w) - \tag{A1-119}$$

$$- aw \frac{\partial}{\partial t}(\delta t) - aw^2 \frac{\partial}{\partial z}(\delta t) - a\delta w - 2k \frac{\partial B}{\partial w} +$$

$$+ k \frac{\partial^2}{\partial u^2}(\delta w) + k \frac{\partial^2}{\partial v^2}(\delta w) + k \frac{\partial^2}{\partial w^2}(\delta w) -$$

$$- u \frac{\partial}{\partial x}(\delta w) - v \frac{\partial}{\partial y}(\delta w) - w \frac{\partial}{\partial z}(\delta w) - \frac{\partial}{\partial t}(\delta w) = 0;$$

$$-ku \frac{\partial}{\partial x}(\delta t) - kv \frac{\partial}{\partial y}(\delta t) - kw \frac{\partial}{\partial z}(\delta t) + 2k \frac{\partial}{\partial u}(\delta u) - k \frac{\partial}{\partial t}(\delta t) = 0; \tag{A1-120}$$

$$-ku \frac{\partial}{\partial x}(\delta t) - kv \frac{\partial}{\partial y}(\delta t) - kw \frac{\partial}{\partial z}(\delta t) + 2k \frac{\partial}{\partial v}(\delta v) - k \frac{\partial}{\partial t}(\delta t) = 0; \tag{A1-121}$$

$$-ku \frac{\partial}{\partial x}(\delta t) - kv \frac{\partial}{\partial y}(\delta t) - kw \frac{\partial}{\partial z}(\delta t) + 2k \frac{\partial}{\partial w}(\delta w) - k \frac{\partial}{\partial t}(\delta t) = 0; \tag{A1-122}$$

$$2k \frac{\partial}{\partial v}(\delta u) + 2k \frac{\partial}{\partial u}(\delta v) = 0; \tag{A1-123}$$

$$2k \frac{\partial}{\partial w}(\delta u) + 2k \frac{\partial}{\partial u}(\delta w) = 0; \tag{A1-124}$$

$$2k \frac{\partial}{\partial w}(\delta v) + 2k \frac{\partial}{\partial v}(\delta w) = 0; \tag{A1-125}$$

$$-au \frac{\partial B}{\partial u} - 3au \frac{\partial}{\partial x}(\delta t) - av \frac{\partial B}{\partial v} - 3av \frac{\partial}{\partial y}(\delta t) - aw \frac{\partial B}{\partial w} - 3aw \frac{\partial}{\partial z}(\delta t) - 3a \frac{\partial}{\partial t}(\delta t) - \tag{A1-126}$$

$$-k \frac{\partial^2 B}{\partial u^2} - k \frac{\partial^2 B}{\partial v^2} - k \frac{\partial^2 B}{\partial w^2} + u \frac{\partial B}{\partial x} + v \frac{\partial B}{\partial y} + w \frac{\partial B}{\partial z} + \frac{\partial B}{\partial t} = 0;$$

$$-au \frac{\partial A}{\partial u} - av \frac{\partial A}{\partial v} - aw \frac{\partial A}{\partial w} - 3aA - \tag{A1-127}$$

$$-k \frac{\partial^2 A}{\partial u^2} - k \frac{\partial^2 A}{\partial v^2} - k \frac{\partial^2 A}{\partial w^2} + u \frac{\partial A}{\partial x} + v \frac{\partial A}{\partial y} + w \frac{\partial A}{\partial z} + \frac{\partial A}{\partial t} = 0.$$

Equation (A1-127) is simply Fokker - Planck equation for $A$.

We solve (A1-114 - A1-116) and find $\delta u, \delta v, \delta w$

$$\delta u = -\left( uv \frac{\partial}{\partial y}(\delta t) + uw \frac{\partial}{\partial z}(\delta t) - u \frac{\partial}{\partial x}(\delta x) + u \frac{\partial}{\partial t}(\delta t) + \tag{A1-128} \right.$$

$$\left. + u^2 \frac{\partial}{\partial x}(\delta t) - v \frac{\partial}{\partial y}(\delta x) - w \frac{\partial}{\partial z}(\delta x) - \frac{\partial}{\partial t}(\delta x) \right\};$$

$$\delta v = -\left( uv \frac{\partial}{\partial x}(\delta t) - u \frac{\partial}{\partial x}(\delta y) + vw \frac{\partial}{\partial z}(\delta t) - v \frac{\partial}{\partial y}(\delta y) + \tag{A1-129} \right.$$

$$\left. + v \frac{\partial}{\partial t}(\delta t) + v^2 \frac{\partial}{\partial y}(\delta t) - w \frac{\partial}{\partial z}(\delta y) - \frac{\partial}{\partial t}(\delta y) \right\};$$

$$\delta w = -\left( uw \frac{\partial}{\partial x}(\delta t) - u \frac{\partial}{\partial x}(\delta z) + vw \frac{\partial}{\partial y}(\delta t) - v \frac{\partial}{\partial y}(\delta z) - \tag{A1-130} \right.$$

$$\left. - w \frac{\partial}{\partial z}(\delta z) + w \frac{\partial}{\partial t}(\delta t) + w^2 \frac{\partial}{\partial z}(\delta t) - \frac{\partial}{\partial t}(\delta z) \right\}.$$

We use these expressions to simplify (A1-117 - A1-126)

$$-2auv \frac{\partial}{\partial y}(\delta t) - 2auw \frac{\partial}{\partial z}(\delta t) - au \frac{\partial}{\partial t}(\delta t) - 2au^2 \frac{\partial}{\partial x}(\delta t) - a \frac{\partial}{\partial t}(\delta x) - 2k \frac{\partial B}{\partial u} - \tag{A1-131}$$

$$-2k \frac{\partial}{\partial x}(\delta t) + 2uvw \frac{\partial^2}{\partial y \partial z}(\delta t) + 2uv \frac{\partial^2}{\partial t \partial y}(\delta t) - 2uv \frac{\partial^2}{\partial x \partial y}(\delta x) + uv^2 \frac{\partial^2}{\partial y^2}(\delta t) + 2uw \frac{\partial^2}{\partial t \partial z}(\delta t) -$$

$$-2uw \frac{\partial^2}{\partial x \partial z}(\delta x) + uw^2 \frac{\partial^2}{\partial z^2}(\delta t) + u \frac{\partial^2}{\partial t^2}(\delta t) - 2u \frac{\partial^2}{\partial t \partial x}(\delta x) + 2u^2 v \frac{\partial^2}{\partial x \partial y}(\delta t)$$

$$+2u^2w\frac{\partial^2}{\partial x\partial z}(\delta t)+2u^2\frac{\partial^2}{\partial t\partial x}(\delta t)-u^2\frac{\partial^2}{\partial x^2}(\delta x)+u^3\frac{\partial^2}{\partial x^2}(\delta t)-2vw\frac{\partial^2}{\partial y\partial z}(\delta x)-$$

$$-2v\frac{\partial^2}{\partial t\partial y}(\delta x)-v^2\frac{\partial^2}{\partial y^2}(\delta x)-2w\frac{\partial^2}{\partial t\partial z}(\delta x)-w^2\frac{\partial^2}{\partial z^2}(\delta x)-\frac{\partial^2}{\partial t^2}(\delta x)=0;$$

$$-2auv\frac{\partial}{\partial x}(\delta t)-2avw\frac{\partial}{\partial z}(\delta t)-av\frac{\partial}{\partial t}(\delta t)-2av^2\frac{\partial}{\partial y}(\delta t)-a\frac{\partial}{\partial t}(\delta y)- \qquad\text{(A1-132)}$$

$$-2k\frac{\partial B}{\partial v}-2k\frac{\partial}{\partial y}(\delta t)+2uvw\frac{\partial^2}{\partial x\partial z}(\delta t)+2uv\frac{\partial^2}{\partial t\partial x}(\delta t)-2uv\frac{\partial^2}{\partial x\partial y}(\delta y)+$$

$$+uv^2\frac{\partial^2}{\partial x\partial y}(\delta t)+uv^2\frac{\partial^2}{\partial x\partial y}(\delta t)-uw\frac{\partial^2}{\partial x\partial z}(\delta y)-uw\frac{\partial^2}{\partial x\partial z}(\delta y)-u\frac{\partial^2}{\partial t\partial x}(\delta y)-u\frac{\partial^2}{\partial t\partial x}(\delta y)+$$

$$+u^2v\frac{\partial^2}{\partial x^2}(\delta t)-u^2\frac{\partial^2}{\partial x^2}(\delta y)+vw\frac{\partial^2}{\partial t\partial z}(\delta t)+vw\frac{\partial^2}{\partial t\partial z}(\delta t)-vw\frac{\partial^2}{\partial y\partial z}(\delta y)-vw\frac{\partial^2}{\partial y\partial z}(\delta y)+$$

$$+vw^2\frac{\partial^2}{\partial z^2}(\delta t)+v\frac{\partial^2}{\partial t^2}(\delta t)-v\frac{\partial^2}{\partial t\partial y}(\delta y)-v\frac{\partial^2}{\partial t\partial y}(\delta y)+v^2w\frac{\partial^2}{\partial y\partial z}(\delta t)+v^2w\frac{\partial^2}{\partial y\partial z}(\delta t)+$$

$$+v^2\frac{\partial^2}{\partial t\partial y}(\delta t)+v^2\frac{\partial^2}{\partial t\partial y}(\delta t)-v^2\frac{\partial^2}{\partial y^2}(\delta y)+v^3\frac{\partial^2}{\partial y^2}(\delta t)-w\frac{\partial^2}{\partial t\partial z}(\delta y)-w\frac{\partial^2}{\partial t\partial z}(\delta y)-w^2\frac{\partial^2}{\partial z^2}(\delta y)-\frac{\partial^2}{\partial t^2}(\delta y)=0;$$

$$-2auw\frac{\partial}{\partial x}(\delta t)-2avw\frac{\partial}{\partial y}(\delta t)-aw\frac{\partial}{\partial t}(\delta t)-2aw^2\frac{\partial}{\partial z}(\delta t)-a\frac{\partial}{\partial t}(\delta z)-2k\frac{\partial B}{\partial w}- \qquad\text{(A1-133)}$$

$$-2k\frac{\partial}{\partial z}(\delta t)+2uvw\frac{\partial^2}{\partial x\partial y}(\delta t)-2uv\frac{\partial^2}{\partial x\partial y}(\delta z)+2uw\frac{\partial^2}{\partial t\partial x}(\delta t)-2uw\frac{\partial^2}{\partial x\partial z}(\delta z)+2uw^2\frac{\partial^2}{\partial x\partial z}(\delta t)-$$

$$-2u\frac{\partial^2}{\partial t\partial x}(\delta z)+u^2w\frac{\partial^2}{\partial x^2}(\delta t)-u^2\frac{\partial^2}{\partial x^2}(\delta z)+2vw\frac{\partial^2}{\partial t\partial y}(\delta t)-2vw\frac{\partial^2}{\partial y\partial z}(\delta z)+2vw^2\frac{\partial^2}{\partial y\partial z}(\delta t)-$$

$$-2v\frac{\partial^2}{\partial t\partial y}(\delta z)+v^2w\frac{\partial^2}{\partial y^2}(\delta t)-v^2\frac{\partial^2}{\partial y^2}(\delta z)+w\frac{\partial^2}{\partial t^2}(\delta t)-2w\frac{\partial^2}{\partial t\partial z}(\delta z)+$$

$$+2w^2\frac{\partial^2}{\partial t\partial z}(\delta t)-w^2\frac{\partial^2}{\partial z^2}(\delta z)+w^3\frac{\partial^2}{\partial z^2}(\delta t)-\frac{\partial^2}{\partial t^2}(\delta z)=0;$$

$$-5ku\frac{\partial}{\partial x}(\delta t)-3kv\frac{\partial}{\partial y}(\delta t)-3kw\frac{\partial}{\partial z}(\delta t)+2k\frac{\partial}{\partial x}(\delta x)-3k\frac{\partial}{\partial t}(\delta t)=0; \qquad\text{(A1-134)}$$

$$-3ku\frac{\partial}{\partial x}(\delta t)-5kv\frac{\partial}{\partial y}(\delta t)-3kw\frac{\partial}{\partial z}(\delta t)+2k\frac{\partial}{\partial y}(\delta y)-3k\frac{\partial}{\partial t}(\delta t)=0; \qquad\text{(A1-135)}$$

$$-3ku\frac{\partial}{\partial x}(\delta t)-3kv\frac{\partial}{\partial y}(\delta t)-5kw\frac{\partial}{\partial z}(\delta t)+2k\frac{\partial}{\partial z}(\delta z)-3k\frac{\partial}{\partial t}(\delta t)=0; \qquad\text{(A1-136)}$$

$$-2ku\frac{\partial}{\partial y}(\delta t)-2kv\frac{\partial}{\partial x}(\delta t)+2k\frac{\partial}{\partial y}(\delta x)+2k\frac{\partial}{\partial x}(\delta y)=0; \qquad\text{(A1-137)}$$

$$-2ku\frac{\partial}{\partial z}(\delta t)-2kw\frac{\partial}{\partial x}(\delta t)+2k\frac{\partial}{\partial z}(\delta x)+2k\frac{\partial}{\partial x}(\delta z)=0; \qquad\text{(A1-138)}$$

$$-2kv \frac{\partial}{\partial z} (\delta t) - 2kw \frac{\partial}{\partial y} (\delta t) + 2k \frac{\partial}{\partial z} (\delta y) + 2k \frac{\partial}{\partial y} (\delta z) = 0; \qquad (A1-139)$$

$$-au \frac{\partial B}{\partial u} - 3au \frac{\partial}{\partial x} (\delta t) - av \frac{\partial B}{\partial v} - 3av \frac{\partial}{\partial y} (\delta t) - aw \frac{\partial B}{\partial w} - 3aw \frac{\partial}{\partial z} (\delta t) - \qquad (A1-140)$$

$$-3a \frac{\partial}{\partial t} (\delta t) - k \frac{\partial^2 B}{\partial u^2} - k \frac{\partial^2 B}{\partial v^2} - k \frac{\partial^2 B}{\partial w^2} + u \frac{\partial B}{\partial x} + v \frac{\partial B}{\partial y} + w \frac{\partial B}{\partial z} + \frac{\partial B}{\partial t} = 0.$$

We collect similar terms in (A1-134 - A1-139) by $u, v, w$, equate these terms to zero and so split (A1-134 - A1-139) into twenty equations:

$$-5k \frac{\partial}{\partial x} (\delta t) = 0; \qquad (A1-141)$$

$$-3k \frac{\partial}{\partial y} (\delta t) = 0; \qquad (A1-142)$$

$$-3k \frac{\partial}{\partial z} (\delta t) = 0; \qquad (A1-143)$$

$$2k \frac{\partial}{\partial x} (\delta x) - 3k \frac{\partial}{\partial t} (\delta t) = 0; \qquad (A1-144)$$

$$-3k \frac{\partial}{\partial x} (\delta t) = 0; \qquad (A1-145)$$

$$-5k \frac{\partial}{\partial y} (\delta t) = 0; \qquad (A1-146)$$

$$-3k \frac{\partial}{\partial z} (\delta t) = 0; \qquad (A1-147)$$

$$2k \frac{\partial}{\partial y} (\delta y) - 3k \frac{\partial}{\partial t} (\delta t) = 0; \qquad (A1-148)$$

$$-3k \frac{\partial}{\partial x} (\delta t) = 0; \qquad (A1-149)$$

$$-3k \frac{\partial}{\partial y} (\delta t) = 0; \qquad (A1-150)$$

$$-5k \frac{\partial}{\partial z} (\delta t) = 0; \qquad (A1-151)$$

$$2k \frac{\partial}{\partial z} (\delta z) - 3k \frac{\partial}{\partial t} (\delta t) = 0; \qquad (A1-152)$$

$$-2k \frac{\partial}{\partial y} (\delta t) = 0; \qquad (A1-153)$$

$$-2k \frac{\partial}{\partial x} (\delta t) = 0; \qquad (A1-154)$$

$$2k \frac{\partial}{\partial y} (\delta x) + 2k \frac{\partial}{\partial x} (\delta y) = 0; \qquad (A1-155)$$

$$-2k \frac{\partial}{\partial z} (\delta t) = 0; \tag{A1-156}$$

$$-2k \frac{\partial}{\partial x} (\delta t) = 0; \tag{A1-157}$$

$$2k \frac{\partial}{\partial z} (\delta x) + 2k \frac{\partial}{\partial x} (\delta z) = 0; \tag{A1-158}$$

$$-2k \frac{\partial}{\partial z} (\delta t) = 0; \tag{A1-159}$$

$$-2k \frac{\partial}{\partial y} (\delta t) = 0; \tag{A1-160}$$

$$2k \frac{\partial}{\partial z} (\delta y) + 2k \frac{\partial}{\partial y} (\delta z) = 0. \tag{A1-161}$$

From (A1-141 - A1-143), (A1-145 - A1-147), (A1-149 - A1-151), (A1-153 - A1-154), (A1-156 - A1-157), (A1-159 - A1-160) we see, that $\delta t = \delta t(t)$. These expressions result in further simplifications

$$-au \frac{\partial}{\partial t} (\delta t) - a \frac{\partial}{\partial t} (\delta x) - 2k \frac{\partial B}{\partial u} - 2uv \frac{\partial^2}{\partial x \partial y} (\delta x) - 2uw \frac{\partial^2}{\partial x \partial z} (\delta x) + \tag{A1-162}$$

$$+u \frac{\partial^2}{\partial t^2} (\delta t) - 2u \frac{\partial^2}{\partial t \partial x} (\delta x) - u^2 \frac{\partial^2}{\partial x^2} (\delta x) - 2vw \frac{\partial^2}{\partial y \partial z} (\delta x) - 2v \frac{\partial^2}{\partial t \partial y} (\delta x) -$$

$$-v^2 \frac{\partial^2}{\partial y^2} (\delta x) - 2w \frac{\partial^2}{\partial t \partial z} (\delta x) - w^2 \frac{\partial^2}{\partial z^2} (\delta x) - \frac{\partial^2}{\partial t^2} (\delta x) = 0;$$

$$-av \frac{\partial}{\partial t} (\delta t) - a \frac{\partial}{\partial t} (\delta y) - 2k \frac{\partial B}{\partial v} - 2uv \frac{\partial^2}{\partial x \partial y} (\delta y) - 2uw \frac{\partial^2}{\partial x \partial z} (\delta y) - 2u \frac{\partial^2}{\partial t \partial x} (\delta y) - \tag{A1-163}$$

$$-u^2 \frac{\partial^2}{\partial x^2} (\delta y) - 2vw \frac{\partial^2}{\partial y \partial z} (\delta y) + v \frac{\partial^2}{\partial t^2} (\delta t) - 2v \frac{\partial^2}{\partial t \partial y} (\delta y) - v^2 \frac{\partial^2}{\partial y^2} (\delta y) -$$

$$-2w \frac{\partial^2}{\partial t \partial z} (\delta y) - w^2 \frac{\partial^2}{\partial z^2} (\delta y) - \frac{\partial^2}{\partial t^2} (\delta y) = 0;$$

$$-aw \frac{\partial}{\partial t} (\delta t) - a \frac{\partial}{\partial t} (\delta z) - 2k \frac{\partial B}{\partial w} - 2uv \frac{\partial^2}{\partial x \partial y} (\delta z) - 2uw \frac{\partial^2}{\partial x \partial z} (\delta z) - \tag{A1-164}$$

$$-2u \frac{\partial^2}{\partial t \partial x} (\delta z) - u^2 \frac{\partial^2}{\partial x^2} (\delta z) - 2vw \frac{\partial^2}{\partial y \partial z} (\delta z) - 2v \frac{\partial^2}{\partial t \partial y} (\delta z) - v^2 \frac{\partial^2}{\partial y^2} (\delta z) +$$

$$+w \frac{\partial^2}{\partial t^2} (\delta t) - 2w \frac{\partial^2}{\partial t \partial z} (\delta z) - w^2 \frac{\partial^2}{\partial z^2} (\delta z) - \frac{\partial^2}{\partial t^2} (\delta z) = 0;$$

$$2k \frac{\partial}{\partial x} (\delta x) - 3k \frac{\partial}{\partial t} (\delta t) = 0; \tag{A1-165}$$

$$2k \frac{\partial}{\partial y} (\delta y) - 3k \frac{\partial}{\partial t} (\delta t) = 0; \tag{A1-166}$$

$$2k \frac{\partial}{\partial z} (\delta z) - 3k \frac{\partial}{\partial t} (\delta t) = 0; \tag{A1-167}$$

$$2k \frac{\partial}{\partial y} (\delta x) + 2k \frac{\partial}{\partial x} (\delta y) = 0; \tag{A1-168}$$

$$2k \frac{\partial}{\partial z} (\delta x) + 2k \frac{\partial}{\partial x} (\delta z) = 0; \tag{A1-169}$$

$$2k \frac{\partial}{\partial z} (\delta y) + 2k \frac{\partial}{\partial y} (\delta z) = 0; \tag{A1-170}$$

$$-au \frac{\partial B}{\partial u} - av \frac{\partial B}{\partial v} - aw \frac{\partial B}{\partial w} - 3a \frac{\partial}{\partial t} (\delta t) - k \frac{\partial^2 B}{\partial u^2} - \tag{A1-171}$$

$$-k \frac{\partial^2 B}{\partial v^2} - k \frac{\partial^2 B}{\partial w^2} + u \frac{\partial B}{\partial x} + v \frac{\partial B}{\partial y} + w \frac{\partial B}{\partial z} + \frac{\partial B}{\partial t} = 0.$$

We integrate (A1-164 - A1-167) and obtain

$$\delta x = C + 3/2x \frac{\partial}{\partial t} (\delta t); \tag{A1-172}$$

$$\delta y = D + 3/2y \frac{\partial}{\partial t} (\delta t); \tag{A1-173}$$

$$\delta z = E + 3/2z \frac{\partial}{\partial t} (\delta t); \tag{A1-174}$$

where $C = C(y, z, t)$, $D = D(x, z, t)$, $E(x, y, t)$. We substitute this expression in (A1-162 - A1-164), (A1-168 - A1-171) and obtain

$$-3/2ax \frac{\partial^2}{\partial t^2} (\delta t) - au \frac{\partial}{\partial t} (\delta t) - a \frac{\partial C}{\partial t} - 2k \frac{\partial B}{\partial u} - 3/2x \frac{\partial^3}{\partial t^3} (\delta t) - \tag{A1-175}$$

$$-2u \frac{\partial^2}{\partial t^2} (\delta t) - 2vw \frac{\partial^2}{\partial y \partial z} (\delta x) - 2v \frac{\partial^2 C}{\partial t \partial y} -$$

$$-v^2 \frac{\partial^2 C}{\partial y^2} - 2w \frac{\partial^2 C}{\partial t \partial z} - w^2 \frac{\partial^2 C}{\partial z^2} - \frac{\partial^2 C}{\partial t^2} = 0;$$

$$-3/2ay \frac{\partial^2}{\partial t^2} (\delta t) - av \frac{\partial}{\partial t} (\delta t) - a \frac{\partial D}{\partial t} - 2k \frac{\partial B}{\partial v} - 3/2y \frac{\partial^3}{\partial t^3} (\delta t) - \tag{A1-176}$$

$$-2uw \frac{\partial^2}{\partial x \partial z} (\delta y) - 2u \frac{\partial^2 D}{\partial t \partial x} - u^2 \frac{\partial^2 D}{\partial x^2} -$$

$$-2v \frac{\partial^2}{\partial t^2} (\delta t) - 2w \frac{\partial^2 D}{\partial t \partial z} - w^2 \frac{\partial^2 D}{\partial z^2} - \frac{\partial^2 D}{\partial t^2} = 0;$$

$$-3/2az \frac{\partial^2}{\partial t^2} (\delta t) - aw \frac{\partial}{\partial t} (\delta t) - a \frac{\partial E}{\partial t} - 2k \frac{\partial B}{\partial w} - 3/2z \frac{\partial^3}{\partial t^3} (\delta t) - \tag{A1-177}$$

$$-2uv \frac{\partial^2}{\partial x \partial y} (\delta z) - 2u \frac{\partial^2 E}{\partial t \partial x} - u^2 \frac{\partial^2 E}{\partial x^2} -$$

$$-2v \frac{\partial^2 E}{\partial t \partial y} - v^2 \frac{\partial^2 E}{\partial y^2} - 2w \frac{\partial^2}{\partial t^2} (\delta t) - \frac{\partial^2 E}{\partial t^2} = 0;$$

$$2k\frac{\partial C}{\partial y}+2k\frac{\partial D}{\partial x}=0; \tag{A1-178}$$

$$2k\frac{\partial C}{\partial z}+2k\frac{\partial E}{\partial x}=0; \tag{A1-179}$$

$$2k\frac{\partial D}{\partial z}+2k\frac{\partial E}{\partial y}=0; \tag{A1-180}$$

$$-au\frac{\partial B}{\partial u}-av\frac{\partial B}{\partial v}-aw\frac{\partial B}{\partial w}-3a\frac{\partial}{\partial t}(\delta t)-k\frac{\partial^2 B}{\partial u^2}- \tag{A1-181}$$

$$-k\frac{\partial^2 B}{\partial v^2}-k\frac{\partial^2 B}{\partial w^2}+u\frac{\partial B}{\partial x}+v\frac{\partial B}{\partial y}+w\frac{\partial B}{\partial z}+\frac{\partial B}{\partial t}=0.$$

We have

$$\frac{\partial C}{\partial x}=\frac{\partial D}{\partial y}=\frac{\partial E}{\partial z}=0 \tag{A1-182}$$

and according to (A1-178 - A1-180)

$$\frac{\partial C}{\partial y}+\frac{\partial D}{\partial x}=\frac{\partial C}{\partial z}+\frac{\partial E}{\partial x}=\frac{\partial D}{\partial z}+\frac{\partial E}{\partial y}=0. \tag{A1-183}$$

Let us think, that $C, D, E$ are components of some displacement vector. According to (A1-182 - A1-183) all components of corresponding deformations tensor are equal to zero. Therefore displacement vector is time-dependent rotation-translation, which preserves Euclidean metric.

$$C=S+Qz-Ry; \tag{A1-184}$$

$$D=T+Rx-Pz; \tag{A1-185}$$

$$E=U+Py-Qx; \tag{A1-186}$$

where $P=P(t)$, $Q=Q(t)$, $R=R(t)$, $S=S(t)$, $T=T(t)$, $U=U(t)$.

We substitute $C, D, E$ to (A1-175 - A1-177) and (A1-181) and obtain

$$-3/2ax\frac{\partial^2}{\partial t^2}(\delta t)+ay\frac{\partial R}{\partial t}-az\frac{\partial Q}{\partial t}-au\frac{\partial}{\partial t}(\delta t)--a\frac{\partial S}{\partial t}- \tag{A1-187}$$

$$-2k\frac{\partial B}{\partial u}-3/2x\frac{\partial^3}{\partial t^3}(\delta t)+y\frac{\partial^2 R}{\partial t^2}-z\frac{\partial^2 Q}{\partial t^2}-2u\frac{\partial^2}{\partial t^2}(\delta t)-$$

$$-2vw\frac{\partial^2}{\partial y\partial z}(\delta x)+2v\frac{\partial R}{\partial t}-2w\frac{\partial Q}{\partial t}-\frac{\partial^2 S}{\partial t^2}=0;$$

$$-ax\frac{\partial R}{\partial t}-3/2ay\frac{\partial^2}{\partial t^2}(\delta t)+az\frac{\partial P}{\partial t}-av\frac{\partial}{\partial t}(\delta t)--a\frac{\partial T}{\partial t}- \tag{A1-188}$$

$$-2k\frac{\partial B}{\partial v}-x\frac{\partial^2 R}{\partial t^2}-3/2y\frac{\partial^3}{\partial t^3}(\delta t)+z\frac{\partial^2 P}{\partial t^2}-2uw\frac{\partial^2}{\partial x\partial z}(\delta y)-$$

$$-2u\frac{\partial R}{\partial t}-2v\frac{\partial^2}{\partial t^2}(\delta t)+2w\frac{\partial P}{\partial t}-\frac{\partial^2 T}{\partial t^2}=0;$$

$$ax\frac{\partial Q}{\partial t}-ay\frac{\partial P}{\partial t}-3/2az\frac{\partial^2}{\partial t^2}(\delta t)-aw\frac{\partial}{\partial t}(\delta t)-a\frac{\partial U}{\partial t}- \tag{A1-189}$$

$$-2k \frac{\partial B}{\partial w} + x \frac{\partial^2 Q}{\partial t^2} - y \frac{\partial^2 P}{\partial t^2} - 3/2z \frac{\partial^3}{\partial t^3} (\delta t) - 2uv \frac{\partial^2}{\partial x \partial y} (\delta z) -$$

$$+2u \frac{\partial Q}{\partial t} - 2v \frac{\partial P}{\partial t} - 2w \frac{\partial^2}{\partial t^2} (\delta t) - \frac{\partial^2 U}{\partial t^2} = 0;$$

$$-au \frac{\partial B}{\partial u} - av \frac{\partial B}{\partial v} - aw \frac{\partial B}{\partial w} - 3a \frac{\partial}{\partial t} (\delta t) - \qquad \text{(A1-190)}$$

$$-k \frac{\partial^2 B}{\partial u^2} - k \frac{\partial^2 B}{\partial v^2} - k \frac{\partial^2 B}{\partial w^2} +$$

$$+u \frac{\partial B}{\partial x} + v \frac{\partial B}{\partial y} + w \frac{\partial B}{\partial z} + \frac{\partial B}{\partial t} = 0.$$

We find $\frac{\partial B}{\partial u}, \frac{\partial B}{\partial v}, \frac{\partial B}{\partial w}$ from (187-189):

$$\frac{\partial B}{\partial u} = -3/2ax \frac{\partial^2}{\partial t^2} (\delta t) + ay \frac{\partial R}{\partial t} - az \frac{\partial Q}{\partial t} - au \frac{\partial}{\partial t} (\delta t) - \qquad \text{(A1-191)}$$

$$-a \frac{\partial S}{\partial t} - 2k \frac{\partial B}{\partial u} - 3/2x \frac{\partial^3}{\partial t^3} (\delta t) + y \frac{\partial^2 R}{\partial t^2} - z \frac{\partial^2 Q}{\partial t^2} -$$

$$-2u \frac{\partial^2}{\partial t^2} (\delta t) - 2vw \frac{\partial^2}{\partial y \partial z} (\delta x) + 2v \frac{\partial R}{\partial t} - 2w \frac{\partial Q}{\partial t} - \frac{\partial^2 S}{\partial t^2};$$

$$\frac{\partial B}{\partial v} = -ax \frac{\partial R}{\partial t} - 3/2ay \frac{\partial^2}{\partial t^2} (\delta t) + az \frac{\partial P}{\partial t} - av \frac{\partial}{\partial t} (\delta t) - \qquad \text{(A1-192)}$$

$$-a \frac{\partial T}{\partial t} - 2k \frac{\partial B}{\partial v} - x \frac{\partial^2 R}{\partial t^2} - 3/2y \frac{\partial^3}{\partial t^3} (\delta t) + z \frac{\partial^2 P}{\partial t^2} -$$

$$-2uw \frac{\partial^2}{\partial x \partial z} (\delta y) - 2u \frac{\partial R}{\partial t} - 2v \frac{\partial^2}{\partial t^2} (\delta t) + 2w \frac{\partial P}{\partial t} - \frac{\partial^2 T}{\partial t^2};$$

$$\frac{\partial B}{\partial w} = ax \frac{\partial Q}{\partial t} - ay \frac{\partial P}{\partial t} - 3/2az \frac{\partial^2}{\partial t^2} (\delta t) - aw \frac{\partial}{\partial t} (\delta t) - a \frac{\partial U}{\partial t} - \qquad \text{(A1-193)}$$

$$-2k \frac{\partial B}{\partial w} + x \frac{\partial^2 Q}{\partial t^2} - y \frac{\partial^2 P}{\partial t^2} - 3/2z \frac{\partial^3}{\partial t^3} (\delta t) - 2uv \frac{\partial^2}{\partial x \partial y} (\delta z) -$$

$$+2u \frac{\partial Q}{\partial t} - 2v \frac{\partial P}{\partial t} - 2w \frac{\partial^2}{\partial t^2} (\delta t) - \frac{\partial^2 U}{\partial t^2};$$

Differentiating (A1-191 - A1-193) by $u$ we have

$$\frac{\partial^2 B}{\partial u^2} = -a \frac{\partial}{\partial t} (\delta t) - 2 \frac{\partial^2}{\partial t^2} (\delta t); \qquad \text{(A1-194)}$$

$$\frac{\partial^2 B}{\partial v \partial u} = -2w \frac{\partial^2}{\partial x \partial z} (\delta y) - 2 \frac{\partial R}{\partial t}; \qquad \text{(A1-195)}$$

$$\frac{\partial^2 B}{\partial w \partial u} = -2v \frac{\partial^2}{\partial x \partial y} (\delta z) + 2 \frac{\partial Q}{\partial t}. \qquad \text{(A1-196)}$$

Differentiating (A1-191 - A1-193) by $v$ we have

$$\frac{\partial^2 B}{\partial u \partial v} = -2w \frac{\partial^2}{\partial y \partial z}(\delta x) + 2 \frac{\partial R}{\partial t};$$ (A1-197)

$$\frac{\partial^2 B}{\partial v^2} = -a \frac{\partial}{\partial t}(\delta t) - 2 \frac{\partial^2}{\partial t^2}(\delta t);$$ (A1-198)

$$\frac{\partial^2 B}{\partial w \partial v} = -2u \frac{\partial^2}{\partial x \partial y}(\delta z) - 2 \frac{\partial P}{\partial t}.$$ (A1-199)

Differentiating (A1-191 - A1-193) by $w$ we have

$$\frac{\partial^2 B}{\partial u \partial w} = -2v \frac{\partial^2}{\partial y \partial z}(\delta x) - 2 \frac{\partial Q}{\partial t};$$ (A1-200)

$$\frac{\partial^2 B}{\partial v \partial w} = -2u \frac{\partial^2}{\partial x \partial z}(\delta y) + 2 \frac{\partial P}{\partial t};$$ (A1-201)

$$\frac{\partial^2 B}{\partial w^2} = -a \frac{\partial}{\partial t}(\delta t) - 2 \frac{\partial^2}{\partial t^2}(\delta t).$$ (A1-202)

We conclude from (194 - 202), that

$$\frac{\partial P}{\partial t} = \frac{\partial Q}{\partial t} = \frac{\partial R}{\partial t} = 0;$$ (A1-203)

$$\frac{\partial^2}{\partial y \partial z}(\delta x) = \frac{\partial^2}{\partial x \partial y}(\delta z) = \frac{\partial^2}{\partial x \partial z}(\delta y) = F.$$ (A1-204)

where $F = F(x, y, z, t)$.

Under these conditions we can integrate (A1-191 - A1-193) and obtain

$$B = \frac{1}{2k} \left( \frac{1}{2}(-a \frac{\partial}{\partial t}(\delta t) - 2 \frac{\partial^2}{\partial t^2}(\delta t))(u^2 + v^2 + w^2) - 2Fuvw + \right.$$ (A1-205)

$$+u(-3/2ax \frac{\partial^2}{\partial t^2}(\delta t) - a \frac{\partial S}{\partial t} - 3/2x \frac{\partial^3}{\partial t^3}(\delta t) - \frac{\partial^2 S}{\partial t^2}) +$$

$$+v(-3/2ay \frac{\partial^2}{\partial t^2}(\delta t) - a \frac{\partial T}{\partial t} - 3/2y \frac{\partial^3}{\partial t^3}(\delta t) - \frac{\partial^2 T}{\partial t^2}) +$$

$$\left. +w(-3/2az \frac{\partial^2}{\partial t^2}(\delta t) - a \frac{\partial U}{\partial t} - 3/2z \frac{\partial^3}{\partial t^3}(\delta t) - \frac{\partial^2 U}{\partial t^2}) \right) + G;$$

where $G = G(x, y, z, t)$.

Substitution of (A1-205) to (A1-171), collecting and equating to zero similar terms by $u, v$ gives

$$3ak^{-1}F - k^{-1} \frac{\partial F}{\partial t} = 0;$$ (A1-206)

$$-k^{-1} \frac{\partial F}{\partial z} = 0;$$ (A1-207)

$$-k^{-1} \frac{\partial F}{\partial y} = 0;$$ (A1-208)

$$3/4a^2k^{-1}x\frac{\partial^2}{\partial t^2}(\delta t) + 1/2a^2k^{-1}\frac{\partial S}{\partial t} - 3/4k^{-1}x\frac{\partial^4}{\partial t^4}(\delta t) - 1/2k^{-1}\frac{\partial^4 T}{\partial t^3\partial x} + \frac{\partial G}{\partial x} = 0; \qquad \text{(A1-209)}$$

$$-k^{-1}\frac{\partial F}{\partial x} = 0; \qquad \text{(A1-210)}$$

$$1/2a^2k^{-1}\frac{\partial}{\partial t}(\delta t) - 5/4k^{-1}\frac{\partial^3}{\partial t^3}(\delta t) = 0; \qquad \text{(A1-211)}$$

$$3/4a^2k^{-1}y\frac{\partial^2}{\partial t^2}(\delta t) + 1/2a^2k^{-1}\frac{\partial T}{\partial t} - 3/4k^{-1}y\frac{\partial^4}{\partial t^4}(\delta t) - 1/2k^{-1}\frac{\partial^4 T}{\partial t^3\partial y} + \frac{\partial G}{\partial y} = 0; \qquad \text{(A1-212)}$$

$$1/2a^2k^{-1}\frac{\partial}{\partial t}(\delta t) - 5/4k^{-1}\frac{\partial^3}{\partial t^3}(\delta t) = 0; \qquad \text{(A1-213)}$$

$$3/4a^2k^{-1}z\frac{\partial^2}{\partial t^2}(\delta t) + 1/2a^2k^{-1}\frac{\partial U}{\partial t} - 3/4k^{-1}z\frac{\partial^4}{\partial t^4}(\delta t) - 1/2k^{-1}\frac{\partial^4 T}{\partial t^3\partial z} + \frac{\partial G}{\partial z} = 0; \qquad \text{(A1-214)}$$

$$1/2a^2k^{-1}\frac{\partial}{\partial t}(\delta t) - 5/4k^{-1}\frac{\partial^3}{\partial t^3}(\delta t) = 0; \qquad \text{(A1-215)}$$

$$-3/2a\frac{\partial}{\partial t}(\delta t) + 3\frac{\partial^2}{\partial t^2}(\delta t) + \frac{\partial G}{\partial t} = 0; \qquad \text{(A1-216)}$$

It follows from (A1-203, A1-207, A1-208, A1-210), that $P, Q, R, F$ are constant

$$P = C_1; \; Q = C_2; \; R = C_3; \; F = C_4 e^{3at}. \qquad \text{(A1-217)}$$

Substitution of (A1-217) to (A1-206 - A1-216) gives

$$3ak^{-1}C_4 = 0; \qquad \text{(A1-218)}$$

$$3/4a^2k^{-1}x\frac{\partial^2}{\partial t^2}(\delta t) + 1/2a^2k^{-1}\frac{\partial S}{\partial t} - 3/4k^{-1}x\frac{\partial^4}{\partial t^4}(\delta t) - 1/2k^{-1}\frac{\partial^3 S}{\partial t^3} + \frac{\partial G}{\partial x} = 0; \qquad \text{(A1-219)}$$

$$1/2a^2k^{-1}\frac{\partial}{\partial t}(\delta t) - 5/4k^{-1}\frac{\partial^3}{\partial t^3}(\delta t) = 0; \qquad \text{(A1-220)}$$

$$3/4a^2k^{-1}y\frac{\partial^2}{\partial t^2}(\delta t) + 1/2a^2k^{-1}\frac{\partial T}{\partial t} - 3/4k^{-1}y\frac{\partial^4}{\partial t^4}(\delta t) - 1/2k^{-1}\frac{\partial^3 T}{\partial t^3} + \frac{\partial G}{\partial y} = 0; \qquad \text{(A1-221)}$$

$$1/2a^2k^{-1}\frac{\partial}{\partial t}(\delta t) - 5/4k^{-1}\frac{\partial^3}{\partial t^3}(\delta t) = 0; \qquad \text{(A1-222)}$$

$$3/4a^2k^{-1}z\frac{\partial^2}{\partial t^2}(\delta t) + 1/2a^2k^{-1}\frac{\partial U}{\partial t} - 3/4k^{-1}z\frac{\partial^4}{\partial t^4}(\delta t) - 1/2k^{-1}\frac{\partial^3 U}{\partial t^3} + \frac{\partial G}{\partial z} = 0; \qquad \text{(A1-223)}$$

$$1/2a^2k^{-1}\frac{\partial}{\partial t}(\delta t) - 5/4k^{-1}\frac{\partial^3}{\partial t^3}(\delta t) = 0; \qquad \text{(A1-224)}$$

$$-3/2a\frac{\partial}{\partial t}(\delta t) + 3\frac{\partial^2}{\partial t^2}(\delta t) + \frac{\partial G}{\partial t} = 0; \qquad \text{(A1-225)}$$

We find derivatives of $G$ from (A1-218), (A1-220), (A1-222), (A1-224)

$$\frac{\partial G}{\partial x} = -(3/4a^2k^{-1}x\frac{\partial^2}{\partial t^2}(\delta t) + 1/2a^2k^{-1}\frac{\partial S}{\partial t} - 3/4k^{-1}x\frac{\partial^4}{\partial t^4}(\delta t) - 1/2k^{-1}\frac{\partial^3 S}{\partial t^3}); \qquad \text{(A1-226)}$$

$$\frac{\partial G}{\partial y} = -(3/4a^2k^{-1}y\frac{\partial^2}{\partial t^2}(\delta t) + 1/2a^2k^{-1}\frac{\partial T}{\partial t} - 3/4k^{-1}y\frac{\partial^4}{\partial t^4}(\delta t) - 1/2k^{-1}\frac{\partial^3 T}{\partial t^3}); \qquad (A1-227)$$

$$\frac{\partial G}{\partial z} = -(3/4a^2k^{-1}z\frac{\partial^2}{\partial t^2}(\delta t) + 1/2a^2k^{-1}\frac{\partial U}{\partial t} - 3/4k^{-1}z\frac{\partial^4}{\partial t^4}(\delta t) - 1/2k^{-1}\frac{\partial^3 U}{\partial t^3}); \qquad (A1-228)$$

$$\frac{\partial G}{\partial t} = -(-3/2a\frac{\partial}{\partial t}(\delta t) + 3\frac{\partial^2}{\partial t^2}(\delta t)); \qquad (A1-229)$$

Cross differentiation of $G$ by $t$ and $x, y, z$ gives

$$\frac{\partial^2 G}{\partial t \partial x} = -(3/4a^2k^{-1}x\frac{\partial^3}{\partial t^3}(\delta t) + 1/2a^2k^{-1}\frac{\partial^2 S}{\partial t^2} - 3/4k^{-1}x\frac{\partial^5}{\partial t^5}(\delta t) - 1/2k^{-1}\frac{\partial^4 S}{\partial t^4}) = \frac{\partial^2 G}{\partial t \partial x} = 0; \quad (A1-230)$$

$$\frac{\partial^2 G}{\partial t \partial y} = -(3/4a^2k^{-1}y\frac{\partial^3}{\partial t^3}(\delta t) + 1/2a^2k^{-1}\frac{\partial^2 T}{\partial t^2} - 3/4k^{-1}y\frac{\partial^5}{\partial t^5}(\delta t) - 1/2k^{-1}\frac{\partial^4 T}{\partial t^4}) = \frac{\partial^2 G}{\partial t \partial y} = 0; \quad (A1-231)$$

$$\frac{\partial^2 G}{\partial t \partial z} = -(3/4a^2k^{-1}z\frac{\partial^3}{\partial t^3}(\delta t) + 1/2a^2k^{-1}\frac{\partial^2 U}{\partial t^2} - 3/4k^{-1}z\frac{\partial^5}{\partial t^5}(\delta t) - 1/2k^{-1}\frac{\partial^4 U}{\partial t^4}) = \frac{\partial^2 G}{\partial t \partial z} = 0; \quad (A1-232)$$

Collecting and equating to zero terms by $x, y, z$ in (A1-230 - A1-232) gives

$$+1/2a^2k^{-1}\frac{\partial^2 S}{\partial t^2} - 1/2k^{-1}\frac{\partial^4 S}{\partial t^4} = 0; \qquad (A1-233)$$

$$+1/2a^2k^{-1}\frac{\partial^2 T}{\partial t^2} - 1/2k^{-1}\frac{\partial^4 T}{\partial t^4} = 0; \qquad (A1-234)$$

$$+1/2a^2k^{-1}\frac{\partial^2 U}{\partial t^2} - 1/2k^{-1}\frac{\partial^4 U}{\partial t^4} = 0; \qquad (A1-235)$$

$$3/4a^2k^{-1}\frac{\partial^3}{\partial t^3}(\delta t) - 3/4k^{-1}\frac{\partial^5}{\partial t^5}(\delta t) = 0. \qquad (A1-236)$$

From (A1-236), (A1-220), (A1-222), (A1-224) we conclude, that

$$\delta t = const = C_5. \qquad (A1-237)$$

Integrating (A1-233 - A1-235) we have

$$S = C_6 + C_7 t + C_8 e^{-at} + C_9 e^{at}; \qquad (A1-238)$$

$$T = C_{10} + C_{11}t + C_{12}e^{-at} + C_{13}e^{at}; \qquad (A1-239)$$

$$U = C_{14} + C_{15}t + C_{16}e^{-at} + C_{17}e^{at}. \qquad (A1-240)$$

It follows, that

$$G = C_{18} - \frac{a^2}{2k}x\,C_{07} - \frac{a^2}{2k}y\,C_{11} - \frac{a^2}{2k}z\,C_{15}. \qquad (A1-241)$$

From (A1-184 - A1-186) we have

$$C = -yC_3 + zC_2 + tC_7 + C_6 + C_8e^{-at} + C_9e^{at}; \qquad (A1-242)$$

$$D = xC_3 - zC_1 + tC_{11} + C_{10} + C_{12}e^{-at} + C_{13}e^{at}; \qquad (A1-243)$$

$$E = -xC_2 + yC_1 + tC_{15} + C_{14} + C_{16}e^{-at} + C_{17}e^{at}. \qquad (A1-244)$$

From (A1-172 - A1-174) we find variations of independent variables

$$\delta x = -yC_3 + zC_2 + tC_7 + C_6 + C_8 e^{-at} + C_9 e^{at};$$ (A1-245)

$$\delta y = xC_3 - zC_1 + tC_{11} + C_{10} + C_{12} e^{-at} + C_{13} e^{at};$$ (A1-246)

$$\delta z = -xC_2 + yC_1 + tC_{15} + C_{14} + C_{16} e^{-at} + C_{17} e^{at};$$ (A1-247)

Cross-differentiating (A1-245 - A1-247), we have (see (A1-204))

$$F = 0.$$

Variations of velocities are equal (see A1-128 - A1-130)

$$\delta u = -aC_8 e^{-at} + aC_9 e^{at} - vC_3 + wC_2 + C_7;$$ (A1-248)

$$\delta v = -aC_{12} e^{-at} + aC_{13} e^{at} + uC_3 - wC_1 + C_{11};$$ (A1-249)

$$\delta w = -aC_{16} e^{-at} + aC_{17} e^{at} - uC_2 + vC_1 + C_{15}.$$ (A1-250)

Variation of $n$ according to (A1-113) is equal to

$$\delta n = -\frac{an}{2k}(uC_7 + vC_{11} + wC_{15}) - \frac{a^2 n}{2k}(xC_7 + yC_{11} + zC_{15}) -$$ (A1-251)

$$-e^{at}\frac{a^2 n}{k}(+uC_9 + vC_{13} + wC_{17}) + nC_{18} + A.$$

This is the desired result.

APPENDIX 2

According to [2] (APPENDIX 1, eq. (A1-8) and (A1-18)), we have for variations of derivatives following expression:

$$\delta \frac{\partial n}{\partial x} = \frac{\partial}{\partial x}(\delta n) + \frac{\partial}{\partial n}(\delta n)\frac{\partial n}{\partial x} - \frac{\partial n}{\partial x}(\frac{\partial}{\partial x}(\delta x) + \frac{\partial}{\partial n}(\delta x)\frac{\partial n}{\partial x}) - \frac{\partial n}{\partial y}(\frac{\partial}{\partial x}(\delta y) + \frac{\partial}{\partial n}(\delta y)\frac{\partial n}{\partial x}) - \text{(A2-1)}$$

$$- \frac{\partial n}{\partial z}(\frac{\partial}{\partial x}(\delta z) + \frac{\partial}{\partial n}(\delta z)\frac{\partial n}{\partial x}) - \frac{\partial n}{\partial u}(\frac{\partial}{\partial x}(\delta u) + \frac{\partial}{\partial n}(\delta u)\frac{\partial n}{\partial x}) - \frac{\partial n}{\partial v}(\frac{\partial}{\partial x}(\delta v) + \frac{\partial}{\partial n}(\delta v)\frac{\partial n}{\partial x}) -$$

$$- \frac{\partial n}{\partial w}(\frac{\partial}{\partial x}(\delta w) + \frac{\partial}{\partial n}(\delta w)\frac{\partial n}{\partial x}) - \frac{\partial n}{\partial t}(\frac{\partial}{\partial x}(\delta t) + \frac{\partial}{\partial n}(\delta t)\frac{\partial n}{\partial x});$$

$$\delta \frac{\partial n}{\partial y} = \frac{\partial}{\partial y}(\delta n) + \frac{\partial}{\partial n}(\delta n)\frac{\partial n}{\partial y} - \frac{\partial n}{\partial x}(\frac{\partial}{\partial y}(\delta x) + \frac{\partial}{\partial n}(\delta x)\frac{\partial n}{\partial y}) - \frac{\partial n}{\partial y}(\frac{\partial}{\partial y}(\delta y) + \frac{\partial}{\partial n}(\delta y)\frac{\partial n}{\partial y}) - \text{(A2-2)}$$

$$- \frac{\partial n}{\partial z}(\frac{\partial}{\partial y}(\delta z) + \frac{\partial}{\partial n}(\delta z)\frac{\partial n}{\partial y}) - \frac{\partial n}{\partial u}(\frac{\partial}{\partial y}(\delta u) + \frac{\partial}{\partial n}(\delta u)\frac{\partial n}{\partial y}) - \frac{\partial n}{\partial v}(\frac{\partial}{\partial y}(\delta v) + \frac{\partial}{\partial n}(\delta v)\frac{\partial n}{\partial y}) -$$

$$- \frac{\partial n}{\partial w}(\frac{\partial}{\partial y}(\delta w) + \frac{\partial}{\partial n}(\delta w)\frac{\partial n}{\partial y}) - \frac{\partial n}{\partial t}(\frac{\partial}{\partial y}(\delta t) + \frac{\partial}{\partial n}(\delta t)\frac{\partial n}{\partial y});$$

$$\delta \frac{\partial n}{\partial z} = \frac{\partial}{\partial z}(\delta n) + \frac{\partial}{\partial n}(\delta n)\frac{\partial n}{\partial z} - \frac{\partial n}{\partial x}(\frac{\partial}{\partial z}(\delta x) + \frac{\partial}{\partial n}(\delta x)\frac{\partial n}{\partial z}) - \frac{\partial n}{\partial y}(\frac{\partial}{\partial z}(\delta y) + \frac{\partial}{\partial n}(\delta y)\frac{\partial n}{\partial z}) - \text{(A2-3)}$$

$$- \frac{\partial n}{\partial z}(\frac{\partial}{\partial z}(\delta z) + \frac{\partial}{\partial n}(\delta z)\frac{\partial n}{\partial z}) - \frac{\partial n}{\partial u}(\frac{\partial}{\partial z}(\delta u) + \frac{\partial}{\partial n}(\delta u)\frac{\partial n}{\partial z}) - \frac{\partial n}{\partial v}(\frac{\partial}{\partial z}(\delta v) + \frac{\partial}{\partial n}(\delta v)\frac{\partial n}{\partial z}) -$$

$$- \frac{\partial n}{\partial w}(\frac{\partial}{\partial z}(\delta w) + \frac{\partial}{\partial n}(\delta w)\frac{\partial n}{\partial z}) - \frac{\partial n}{\partial t}(\frac{\partial}{\partial z}(\delta t) + \frac{\partial}{\partial n}(\delta t)\frac{\partial n}{\partial z});$$

$$\delta \frac{\partial n}{\partial u} = \frac{\partial}{\partial u}(\delta n) + \frac{\partial}{\partial n}(\delta n)\frac{\partial n}{\partial u} - \frac{\partial n}{\partial x}(\frac{\partial}{\partial u}(\delta x) + \frac{\partial}{\partial n}(\delta x)\frac{\partial n}{\partial u}) - \frac{\partial n}{\partial y}(\frac{\partial}{\partial u}(\delta y) + \frac{\partial}{\partial n}(\delta y)\frac{\partial n}{\partial u}) - \text{(A2-4)}$$

$$- \frac{\partial n}{\partial z}(\frac{\partial}{\partial u}(\delta z) + \frac{\partial}{\partial n}(\delta z)\frac{\partial n}{\partial u}) - \frac{\partial n}{\partial u}(\frac{\partial}{\partial u}(\delta u) + \frac{\partial}{\partial n}(\delta u)\frac{\partial n}{\partial u}) - \frac{\partial n}{\partial v}(\frac{\partial}{\partial u}(\delta v) + \frac{\partial}{\partial n}(\delta v)\frac{\partial n}{\partial u}) -$$

$$- \frac{\partial n}{\partial w}(\frac{\partial}{\partial u}(\delta w) + \frac{\partial}{\partial n}(\delta w)\frac{\partial n}{\partial u}) - \frac{\partial n}{\partial t}(\frac{\partial}{\partial u}(\delta t) + \frac{\partial}{\partial n}(\delta t)\frac{\partial n}{\partial u});$$

$$\delta \frac{\partial n}{\partial v} = \frac{\partial}{\partial v}(\delta n) + \frac{\partial}{\partial n}(\delta n)\frac{\partial n}{\partial v} - \frac{\partial n}{\partial x}(\frac{\partial}{\partial v}(\delta x) + \frac{\partial}{\partial n}(\delta x)\frac{\partial n}{\partial v}) - \frac{\partial n}{\partial y}(\frac{\partial}{\partial v}(\delta y) + \frac{\partial}{\partial n}(\delta y)\frac{\partial n}{\partial v}) - \text{(A2-5)}$$

$$- \frac{\partial n}{\partial z}(\frac{\partial}{\partial v}(\delta z) + \frac{\partial}{\partial n}(\delta z)\frac{\partial n}{\partial v}) - \frac{\partial n}{\partial u}(\frac{\partial}{\partial v}(\delta u) + \frac{\partial}{\partial n}(\delta u)\frac{\partial n}{\partial v}) - \frac{\partial n}{\partial v}(\frac{\partial}{\partial v}(\delta v) + \frac{\partial}{\partial n}(\delta v)\frac{\partial n}{\partial v}) -$$

$$- \frac{\partial n}{\partial w}(\frac{\partial}{\partial v}(\delta w) + \frac{\partial}{\partial n}(\delta w)\frac{\partial n}{\partial v}) - \frac{\partial n}{\partial t}(\frac{\partial}{\partial v}(\delta t) + \frac{\partial}{\partial n}(\delta t)\frac{\partial n}{\partial v});$$

$$\delta \frac{\partial n}{\partial w} = \frac{\partial}{\partial w}(\delta n) + \frac{\partial}{\partial n}(\delta n)\frac{\partial n}{\partial w} - \frac{\partial n}{\partial x}(\frac{\partial}{\partial w}(\delta x) + \frac{\partial}{\partial n}(\delta x)\frac{\partial n}{\partial w}) - \frac{\partial n}{\partial y}(\frac{\partial}{\partial w}(\delta y) + \frac{\partial}{\partial n}(\delta y)\frac{\partial n}{\partial w}) - \text{(A2-6)}$$

$$- \frac{\partial n}{\partial z}(\frac{\partial}{\partial w}(\delta z) + \frac{\partial}{\partial n}(\delta z)\frac{\partial n}{\partial w}) - \frac{\partial n}{\partial u}(\frac{\partial}{\partial w}(\delta u) + \frac{\partial}{\partial n}(\delta u)\frac{\partial n}{\partial w}) - \frac{\partial n}{\partial v}(\frac{\partial}{\partial w}(\delta v) + \frac{\partial}{\partial n}(\delta v)\frac{\partial n}{\partial w}) -$$

$$-\frac{\partial n}{\partial w}(\frac{\partial}{\partial w}(\delta w)+\frac{\partial}{\partial n}(\delta w)\frac{\partial n}{\partial w})-\frac{\partial n}{\partial t}(\frac{\partial}{\partial w}(\delta t)+\frac{\partial}{\partial n}(\delta t)\frac{\partial n}{\partial w});$$

$$\delta\frac{\partial n}{\partial t}=\frac{\partial}{\partial t}(\delta n)+\frac{\partial}{\partial n}(\delta n)\frac{\partial n}{\partial t}-\frac{\partial n}{\partial x}(\frac{\partial}{\partial t}(\delta x)+\frac{\partial}{\partial n}(\delta x)\frac{\partial n}{\partial t})-\frac{\partial n}{\partial y}(\frac{\partial}{\partial t}(\delta y)+\frac{\partial}{\partial n}(\delta y)\frac{\partial n}{\partial t})-\quad\text{(A2-7)}$$

$$-\frac{\partial n}{\partial z}(\frac{\partial}{\partial t}(\delta z)+\frac{\partial}{\partial n}(\delta z)\frac{\partial n}{\partial t})-\frac{\partial n}{\partial u}(\frac{\partial}{\partial t}(\delta u)+\frac{\partial}{\partial n}(\delta u)\frac{\partial n}{\partial t})-\frac{\partial n}{\partial v}(\frac{\partial}{\partial t}(\delta v)+\frac{\partial}{\partial n}(\delta v)\frac{\partial n}{\partial t})-$$

$$-\frac{\partial n}{\partial w}(\frac{\partial}{\partial t}(\delta w)+\frac{\partial}{\partial n}(\delta w)\frac{\partial n}{\partial t})-\frac{\partial n}{\partial t}(\frac{\partial}{\partial t}(\delta t)+\frac{\partial}{\partial n}(\delta t)\frac{\partial n}{\partial t});$$

$$\delta\frac{\partial^2 n}{\partial u^2}=\frac{\partial}{\partial n}(\delta n)\frac{\partial^2 n}{\partial u^2}-\frac{\partial^2 n}{\partial u\partial x}(\frac{\partial}{\partial u}(\delta x)+\frac{\partial}{\partial n}(\delta x)\frac{\partial n}{\partial u})-\frac{\partial^2 n}{\partial u\partial y}(\frac{\partial}{\partial u}(\delta y)+\frac{\partial}{\partial n}(\delta y)\frac{\partial n}{\partial u})-\frac{\partial^2 n}{\partial u\partial z}(\frac{\partial}{\partial u}(\delta z)+\frac{\partial}{\partial n}(\delta z)\frac{\partial n}{\partial u})-\text{(A2-8)}$$

$$-\frac{\partial^2 n}{\partial u^2}(\frac{\partial}{\partial u}(\delta u)+\frac{\partial}{\partial n}(\delta u)\frac{\partial n}{\partial u})-\frac{\partial^2 n}{\partial u\partial v}(\frac{\partial}{\partial u}(\delta v)+\frac{\partial}{\partial n}(\delta v)\frac{\partial n}{\partial u})-\frac{\partial^2 n}{\partial u\partial w}(\frac{\partial}{\partial u}(\delta w)+\frac{\partial}{\partial n}(\delta w)\frac{\partial n}{\partial u})-\frac{\partial^2 n}{\partial t\partial u}(\frac{\partial}{\partial u}(\delta t)+\frac{\partial}{\partial n}(\delta t)\frac{\partial n}{\partial u})-$$

$$-\frac{\partial^2 n}{\partial u^2}\frac{\partial n}{\partial x}\frac{\partial}{\partial n}(\delta x)-\frac{\partial^2 n}{\partial u^2}\frac{\partial n}{\partial y}\frac{\partial}{\partial n}(\delta y)-\frac{\partial^2 n}{\partial u^2}\frac{\partial n}{\partial z}\frac{\partial}{\partial n}(\delta z)-\frac{\partial^2 n}{\partial u^2}\frac{\partial n}{\partial u}\frac{\partial}{\partial n}(\delta u)$$

$$-\frac{\partial^2 n}{\partial u^2}\frac{\partial n}{\partial v}\frac{\partial}{\partial n}(\delta v)-\frac{\partial^2 n}{\partial u^2}\frac{\partial n}{\partial w}\frac{\partial}{\partial n}(\delta w)-\frac{\partial^2 n}{\partial u^2}\frac{\partial n}{\partial t}\frac{\partial}{\partial n}(\delta t)+\frac{\partial^2}{\partial u^2}(\delta n)+$$

$$+\frac{\partial^2}{\partial n\partial u}(\delta n)\frac{\partial n}{\partial u}+\frac{\partial n}{\partial u}(\frac{\partial^2}{\partial n\partial u}(\delta n)+\frac{\partial^2}{\partial n^2}(\delta n)\frac{\partial n}{\partial u})-\frac{\partial n}{\partial x}(\frac{\partial^2}{\partial u^2}(\delta x)+\frac{\partial^2}{\partial n\partial u}(\delta x)\frac{\partial n}{\partial u})-\frac{\partial n}{\partial y}(\frac{\partial^2}{\partial u^2}(\delta y)+\frac{\partial^2}{\partial n\partial u}(\delta y)\frac{\partial n}{\partial u})-$$

$$-\frac{\partial n}{\partial z}(\frac{\partial^2}{\partial u^2}(\delta z)+\frac{\partial^2}{\partial n\partial u}(\delta z)\frac{\partial n}{\partial u})-\frac{\partial n}{\partial u}(\frac{\partial^2}{\partial u^2}(\delta u)+\frac{\partial^2}{\partial n\partial u}(\delta u)\frac{\partial n}{\partial u})-\frac{\partial n}{\partial v}(\frac{\partial^2}{\partial u^2}(\delta v)+\frac{\partial^2}{\partial n\partial u}(\delta v)\frac{\partial n}{\partial u})+$$

$$-\frac{\partial n}{\partial w}(\frac{\partial^2}{\partial u^2}(\delta w)+\frac{\partial^2}{\partial n\partial u}(\delta w)\frac{\partial n}{\partial u})-\frac{\partial n}{\partial t}(\frac{\partial^2}{\partial u^2}(\delta t)+\frac{\partial^2}{\partial n\partial u}(\delta t)\frac{\partial n}{\partial u})-\frac{\partial n}{\partial x}\frac{\partial n}{\partial u}(\frac{\partial^2}{\partial n\partial u}(\delta x)+\frac{\partial^2}{\partial n^2}(\delta x)\frac{\partial n}{\partial u})+$$

$$-\frac{\partial n}{\partial y}\frac{\partial n}{\partial u}(\frac{\partial^2}{\partial n\partial u}(\delta y)+\frac{\partial^2}{\partial n^2}(\delta y)\frac{\partial n}{\partial u})-\frac{\partial n}{\partial z}\frac{\partial n}{\partial u}(\frac{\partial^2}{\partial n\partial u}(\delta z)+\frac{\partial^2}{\partial n^2}(\delta z)\frac{\partial n}{\partial u})-\frac{\partial n}{\partial u}\frac{\partial n}{\partial u}(\frac{\partial^2}{\partial n\partial u}(\delta u)+\frac{\partial^2}{\partial n^2}(\delta u)\frac{\partial n}{\partial u})-$$

$$-\frac{\partial n}{\partial v}\frac{\partial n}{\partial u}(\frac{\partial^2}{\partial n\partial u}(\delta v)+\frac{\partial^2}{\partial n^2}(\delta v)\frac{\partial n}{\partial u})-\frac{\partial n}{\partial w}\frac{\partial n}{\partial u}(\frac{\partial^2}{\partial n\partial u}(\delta w)+\frac{\partial^2}{\partial n^2}(\delta w)\frac{\partial n}{\partial u})-\frac{\partial n}{\partial t}\frac{\partial n}{\partial u}(\frac{\partial^2}{\partial n\partial u}(\delta t)+\frac{\partial^2}{\partial n^2}(\delta t)\frac{\partial n}{\partial u})-$$

$$-\frac{\partial^2 n}{\partial u\partial x}(\frac{\partial}{\partial u}(\delta x)+\frac{\partial}{\partial n}(\delta x)\frac{\partial n}{\partial u})-\frac{\partial^2 n}{\partial u\partial y}(\frac{\partial}{\partial u}(\delta y)+\frac{\partial}{\partial n}(\delta y)\frac{\partial n}{\partial u})-\frac{\partial^2 n}{\partial u\partial z}(\frac{\partial}{\partial u}(\delta z)+\frac{\partial}{\partial n}(\delta z)\frac{\partial n}{\partial u})-$$

$$-\frac{\partial^2 n}{\partial u^2}(\frac{\partial}{\partial u}(\delta u)+\frac{\partial}{\partial n}(\delta u)\frac{\partial n}{\partial u})-\frac{\partial^2 n}{\partial u\partial v}(\frac{\partial}{\partial u}(\delta v)+\frac{\partial}{\partial n}(\delta v)\frac{\partial n}{\partial u})-\frac{\partial^2 n}{\partial u\partial w}(\frac{\partial}{\partial u}(\delta w)+\frac{\partial}{\partial n}(\delta w)\frac{\partial n}{\partial u})-\frac{\partial^2 n}{\partial t\partial u}(\frac{\partial}{\partial u}(\delta t)+\frac{\partial}{\partial n}(\delta t)\frac{\partial n}{\partial u});$$

$$\delta\frac{\partial^2 n}{\partial v^2}=\frac{\partial}{\partial n}(\delta n)\frac{\partial^2 n}{\partial v^2}-\frac{\partial^2 n}{\partial v\partial x}(\frac{\partial}{\partial v}(\delta x)+\frac{\partial}{\partial n}(\delta x)\frac{\partial n}{\partial v})-\frac{\partial^2 n}{\partial v\partial y}(\frac{\partial}{\partial v}(\delta y)+\frac{\partial}{\partial n}(\delta y)\frac{\partial n}{\partial v})-\frac{\partial^2 n}{\partial v\partial z}(\frac{\partial}{\partial v}(\delta z)+\frac{\partial}{\partial n}(\delta z)\frac{\partial n}{\partial v})-\text{(A2-9)}$$

$$-\frac{\partial^2 n}{\partial u\partial v}(\frac{\partial}{\partial v}(\delta u)+\frac{\partial}{\partial n}(\delta u)\frac{\partial n}{\partial v})-\frac{\partial^2 n}{\partial v^2}(\frac{\partial}{\partial v}(\delta v)+\frac{\partial}{\partial n}(\delta v)\frac{\partial n}{\partial v})-\frac{\partial^2 n}{\partial v\partial w}(\frac{\partial}{\partial v}(\delta w)+\frac{\partial}{\partial n}(\delta w)\frac{\partial n}{\partial v})-\frac{\partial^2 n}{\partial t\partial v}(\frac{\partial}{\partial v}(\delta t)+\frac{\partial}{\partial n}(\delta t)\frac{\partial n}{\partial v})-$$

$$-\frac{\partial^2 n}{\partial v^2}\frac{\partial n}{\partial x}\frac{\partial}{\partial n}(\delta x)-\frac{\partial^2 n}{\partial v^2}\frac{\partial n}{\partial y}\frac{\partial}{\partial n}(\delta y)-\frac{\partial^2 n}{\partial v^2}\frac{\partial n}{\partial z}\frac{\partial}{\partial n}(\delta z)-\frac{\partial^2 n}{\partial v^2}\frac{\partial n}{\partial u}\frac{\partial}{\partial n}(\delta u)-$$

$$-\frac{\partial^2 n}{\partial v^2}\frac{\partial n}{\partial v}\frac{\partial}{\partial n}(\delta v)-\frac{\partial^2 n}{\partial v^2}\frac{\partial n}{\partial w}\frac{\partial}{\partial n}(\delta w)-\frac{\partial^2 n}{\partial v^2}\frac{\partial n}{\partial t}\frac{\partial}{\partial n}(\delta t)+\frac{\partial^2}{\partial v^2}(\delta n)+$$

$$+\frac{\partial^2}{\partial n\partial v}(\delta n)\frac{\partial n}{\partial v}+\frac{\partial n}{\partial v}(\frac{\partial^2}{\partial n\partial v}(\delta n)+\frac{\partial^2}{\partial n^2}(\delta n)\frac{\partial n}{\partial v})-\frac{\partial n}{\partial x}(\frac{\partial^2}{\partial v^2}(\delta x)+\frac{\partial^2}{\partial n\partial v}(\delta x)\frac{\partial n}{\partial v})-\frac{\partial n}{\partial y}(\frac{\partial^2}{\partial v^2}(\delta y)+\frac{\partial^2}{\partial n\partial v}(\delta y)\frac{\partial n}{\partial v})-$$

$$-\frac{\partial n}{\partial z}(\frac{\partial^2}{\partial v^2}(\delta z)+\frac{\partial^2}{\partial n\partial v}(\delta z)\frac{\partial n}{\partial v})-\frac{\partial n}{\partial u}(\frac{\partial^2}{\partial v^2}(\delta u)+\frac{\partial^2}{\partial n\partial v}(\delta u)\frac{\partial n}{\partial v})-\frac{\partial n}{\partial v}(\frac{\partial^2}{\partial v^2}(\delta v)+\frac{\partial^2}{\partial n\partial v}(\delta v)\frac{\partial n}{\partial v})-$$

$$-\frac{\partial n}{\partial w}(\frac{\partial^2}{\partial v^2}(\delta w)+\frac{\partial^2}{\partial n\partial v}(\delta w)\frac{\partial n}{\partial v})-\frac{\partial n}{\partial t}(\frac{\partial^2}{\partial v^2}(\delta t)+\frac{\partial^2}{\partial n\partial v}(\delta t)\frac{\partial n}{\partial v})-\frac{\partial n}{\partial x}\frac{\partial n}{\partial v}(\frac{\partial^2}{\partial n\partial v}(\delta x)+\frac{\partial^2}{\partial n^2}(\delta x)\frac{\partial n}{\partial v})-$$

$$-\frac{\partial n}{\partial y}\frac{\partial n}{\partial v}(\frac{\partial^2}{\partial n\partial v}(\delta y)+\frac{\partial^2}{\partial n^2}(\delta y)\frac{\partial n}{\partial v})-\frac{\partial n}{\partial z}\frac{\partial n}{\partial v}(\frac{\partial^2}{\partial n\partial v}(\delta z)+\frac{\partial^2}{\partial n^2}(\delta z)\frac{\partial n}{\partial v})-\frac{\partial n}{\partial u}\frac{\partial n}{\partial v}(\frac{\partial^2}{\partial n\partial v}(\delta u)+\frac{\partial^2}{\partial n^2}(\delta u)\frac{\partial n}{\partial v})-$$

$$-\frac{\partial n}{\partial v}\frac{\partial n}{\partial v}(\frac{\partial^2}{\partial n\partial v}(\delta v)+\frac{\partial^2}{\partial n^2}(\delta v)\frac{\partial n}{\partial v})-\frac{\partial n}{\partial w}\frac{\partial n}{\partial v}(\frac{\partial^2}{\partial n\partial v}(\delta w)+\frac{\partial^2}{\partial n^2}(\delta w)\frac{\partial n}{\partial v})-\frac{\partial n}{\partial t}\frac{\partial n}{\partial v}(\frac{\partial^2}{\partial n\partial v}(\delta t)+\frac{\partial^2}{\partial n^2}(\delta t)\frac{\partial n}{\partial v})-$$

$$-\frac{\partial^2 n}{\partial v\partial x}(\frac{\partial}{\partial v}(\delta x)+\frac{\partial}{\partial n}(\delta x)\frac{\partial n}{\partial v})-\frac{\partial^2 n}{\partial v\partial y}(\frac{\partial}{\partial v}(\delta y)+\frac{\partial}{\partial n}(\delta y)\frac{\partial n}{\partial v})-\frac{\partial^2 n}{\partial v\partial z}(\frac{\partial}{\partial v}(\delta z)+\frac{\partial}{\partial n}(\delta z)\frac{\partial n}{\partial v})-$$

$$-\frac{\partial^2 n}{\partial u\partial v}(\frac{\partial}{\partial v}(\delta u)+\frac{\partial}{\partial n}(\delta u)\frac{\partial n}{\partial v})-\frac{\partial^2 n}{\partial v^2}(\frac{\partial}{\partial v}(\delta v)+\frac{\partial}{\partial n}(\delta v)\frac{\partial n}{\partial v})-\frac{\partial^2 n}{\partial v\partial w}(\frac{\partial}{\partial v}(\delta w)+\frac{\partial}{\partial n}(\delta w)\frac{\partial n}{\partial v})-\frac{\partial^2 n}{\partial t\partial v}(\frac{\partial}{\partial v}(\delta t)+\frac{\partial}{\partial n}(\delta t)\frac{\partial n}{\partial v});$$

$$\delta\frac{\partial^2 n}{\partial w^2}=\frac{\partial}{\partial n}(\delta n)\frac{\partial^2 n}{\partial w^2}-\frac{\partial^2 n}{\partial w\partial x}(\frac{\partial}{\partial w}(\delta x)+\frac{\partial}{\partial n}(\delta x)\frac{\partial n}{\partial w})-\frac{\partial^2 n}{\partial w\partial y}(\frac{\partial}{\partial w}(\delta y)+\frac{\partial}{\partial n}(\delta y)\frac{\partial n}{\partial w})-\text{(A2-10)}$$

$$-\frac{\partial^2 n}{\partial w\partial z}(\frac{\partial}{\partial w}(\delta z)+\frac{\partial}{\partial n}(\delta z)\frac{\partial n}{\partial w})-\frac{\partial^2 n}{\partial u\partial w}(\frac{\partial}{\partial w}(\delta u)+\frac{\partial}{\partial n}(\delta u)\frac{\partial n}{\partial w})-\frac{\partial^2 n}{\partial v\partial w}(\frac{\partial}{\partial w}(\delta v)+\frac{\partial}{\partial n}(\delta v)\frac{\partial n}{\partial w})-$$

$$-\frac{\partial^2 n}{\partial w^2}(\frac{\partial}{\partial w}(\delta w)+\frac{\partial}{\partial n}(\delta w)\frac{\partial n}{\partial w})-\frac{\partial^2 n}{\partial t\partial w}(\frac{\partial}{\partial w}(\delta t)+\frac{\partial}{\partial n}(\delta t)\frac{\partial n}{\partial w})-$$

$$-\frac{\partial^2 n}{\partial w^2}\frac{\partial n}{\partial x}\frac{\partial}{\partial n}(\delta x)-\frac{\partial^2 n}{\partial w^2}\frac{\partial n}{\partial y}\frac{\partial}{\partial n}(\delta y)-\frac{\partial^2 n}{\partial w^2}\frac{\partial n}{\partial z}\frac{\partial}{\partial n}(\delta z)-\frac{\partial^2 n}{\partial w^2}\frac{\partial n}{\partial u}\frac{\partial}{\partial n}(\delta u)-$$

$$-\frac{\partial^2 n}{\partial w^2}\frac{\partial n}{\partial v}\frac{\partial}{\partial n}(\delta v)-\frac{\partial^2 n}{\partial w^2}\frac{\partial n}{\partial w}\frac{\partial}{\partial n}(\delta w)-\frac{\partial^2 n}{\partial w^2}\frac{\partial n}{\partial t}\frac{\partial}{\partial n}(\delta t)+\frac{\partial^2}{\partial w^2}(\delta n)+$$

$$+\frac{\partial^2}{\partial n\partial w}(\delta n)\frac{\partial n}{\partial w}+\frac{\partial n}{\partial w}(\frac{\partial^2}{\partial n\partial w}(\delta n)+\frac{\partial^2}{\partial n^2}(\delta n)\frac{\partial n}{\partial w})-\frac{\partial n}{\partial x}(\frac{\partial^2}{\partial w^2}(\delta x)+\frac{\partial^2}{\partial n\partial w}(\delta x)\frac{\partial n}{\partial w})-$$

$$-\frac{\partial n}{\partial y}(\frac{\partial^2}{\partial w^2}(\delta y)+\frac{\partial^2}{\partial n\partial w}(\delta y)\frac{\partial n}{\partial w})-\frac{\partial n}{\partial z}(\frac{\partial^2}{\partial w^2}(\delta z)+\frac{\partial^2}{\partial n\partial w}(\delta z)\frac{\partial n}{\partial w})-\frac{\partial n}{\partial u}(\frac{\partial^2}{\partial w^2}(\delta u)+\frac{\partial^2}{\partial n\partial w}(\delta u)\frac{\partial n}{\partial w})-$$

$$-\frac{\partial n}{\partial v}(\frac{\partial^2}{\partial w^2}(\delta v)+\frac{\partial^2}{\partial n\partial w}(\delta v)\frac{\partial n}{\partial w})-\frac{\partial n}{\partial w}(\frac{\partial^2}{\partial w^2}(\delta w)+\frac{\partial^2}{\partial n\partial w}(\delta w)\frac{\partial n}{\partial w})-\frac{\partial n}{\partial t}(\frac{\partial^2}{\partial w^2}(\delta t)+\frac{\partial^2}{\partial n\partial w}(\delta t)\frac{\partial n}{\partial w})-$$

$$-\frac{\partial n}{\partial x}\frac{\partial n}{\partial w}(\frac{\partial^2}{\partial n\partial w}(\delta x)+\frac{\partial^2}{\partial n^2}(\delta x)\frac{\partial n}{\partial w})-\frac{\partial n}{\partial y}\frac{\partial n}{\partial w}(\frac{\partial^2}{\partial n\partial w}(\delta y)+\frac{\partial^2}{\partial n^2}(\delta y)\frac{\partial n}{\partial w})-\frac{\partial n}{\partial z}\frac{\partial n}{\partial w}(\frac{\partial^2}{\partial n\partial w}(\delta z)+\frac{\partial^2}{\partial n^2}(\delta z)\frac{\partial n}{\partial w})-$$

$$-\frac{\partial n}{\partial u}\frac{\partial n}{\partial w}(\frac{\partial^2}{\partial n\partial w}(\delta u)+\frac{\partial^2}{\partial n^2}(\delta u)\frac{\partial n}{\partial w})-\frac{\partial n}{\partial v}\frac{\partial n}{\partial w}(\frac{\partial^2}{\partial n\partial w}(\delta v)+\frac{\partial^2}{\partial n^2}(\delta v)\frac{\partial n}{\partial w})-\frac{\partial n}{\partial w}\frac{\partial n}{\partial w}(\frac{\partial^2}{\partial n\partial w}(\delta w)+\frac{\partial^2}{\partial n^2}(\delta w)\frac{\partial n}{\partial w})-$$

$$-\frac{\partial n}{\partial t}\frac{\partial n}{\partial w}\left(\frac{\partial^2}{\partial n\partial w}\left(\delta t\right)+\frac{\partial^2}{\partial n^2}\left(\delta t\right)\frac{\partial n}{\partial w}\right)-\frac{\partial^2 n}{\partial w\partial x}\left(\frac{\partial}{\partial w}\left(\delta x\right)+\frac{\partial}{\partial n}\left(\delta x\right)\frac{\partial n}{\partial w}\right)-\frac{\partial^2 n}{\partial w\partial y}\left(\frac{\partial}{\partial w}\left(\delta y\right)+\frac{\partial}{\partial n}\left(\delta y\right)\frac{\partial n}{\partial w}\right)-$$

$$-\frac{\partial^2 n}{\partial w\partial z}\left(\frac{\partial}{\partial w}\left(\delta z\right)+\frac{\partial}{\partial n}\left(\delta z\right)\frac{\partial n}{\partial w}\right)-\frac{\partial^2 n}{\partial u\partial w}\left(\frac{\partial}{\partial w}\left(\delta u\right)+\frac{\partial}{\partial n}\left(\delta u\right)\frac{\partial n}{\partial w}\right)-\frac{\partial^2 n}{\partial v\partial w}\left(\frac{\partial}{\partial w}\left(\delta v\right)+\frac{\partial}{\partial n}\left(\delta v\right)\frac{\partial n}{\partial w}\right)-$$

$$-\frac{\partial^2 n}{\partial w^2}\left(\frac{\partial}{\partial w}\left(\delta w\right)+\frac{\partial}{\partial n}\left(\delta w\right)\frac{\partial n}{\partial w}\right)-\frac{\partial^2 n}{\partial t\partial w}\left(\frac{\partial}{\partial w}\left(\delta t\right)+\frac{\partial}{\partial n}\left(\delta t\right)\frac{\partial n}{\partial w}\right);$$